\documentclass[sn-mathphys,Numbered]{sn-jnl}

\usepackage{graphicx}%
\usepackage{multirow}%
\usepackage{amsmath,amssymb,amsfonts}%
\usepackage{amsthm}%
\usepackage{mathrsfs}%
\usepackage[title]{appendix}%
\usepackage{xcolor}%
\usepackage{textcomp}%
\usepackage{manyfoot}%
\usepackage{booktabs}%
\usepackage{algorithm}%
\usepackage{algorithmicx}%
\usepackage{algpseudocode}%
\usepackage{listings}%

\usepackage{tabularx}
\usepackage{mathtools}
\usepackage{bm}
\usepackage{array}
\usepackage{subfig}
\usepackage{stfloats}
\usepackage{color}
\usepackage{mathrsfs}
\usepackage{hhline}
\usepackage{url} %
\usepackage{setspace}
\usepackage{morefloats} %
\usepackage{float}

\theoremstyle{thmstyleone}%
\theoremstyle{thmstyletwo}%

\theoremstyle{thmstylethree}%

\begin{document}

\title[Article Title]{Transformer-based Autoencoder with ID Constraint for Unsupervised Anomalous Sound Detection}

\author*[1]{\fnm{Jian} \sur{Guan}}\email{j.guan@hrbeu.edu.cn}

\author[1]{\fnm{Youde} \sur{Liu}}\email{liuyoude@hrbeu.edu.cn}

\author[2]{\fnm{Qiuqiang} \sur{Kong}}\email{qiuqiangkong@gmail.com}

\author[1]{\fnm{Feiyang} \sur{Xiao}}\email{xiaofeiyang128@gmail.com}

\author[3]{\fnm{Qiaoxi} \sur{Zhu}}\email{qiaoxi.zhu@gmail.com}

\author[1]{\fnm{Jiantong} \sur{Tian}}\email{tianjiantong2022@hrbeu.edu.cn}

\author[4]{\fnm{Wenwu} \sur{Wang}}\email{w.wang@surrey.ac.uk}

\affil*[1]{\orgdiv{College of Computer Science and Technology}, \orgname{Harbin Engineering University}, \orgaddress{\city{Harbin}, \postcode{150001}, \state{Heilongjiang}, \country{China}}}

\affil[2]{\orgname{Bytedance}, \orgaddress{\city{Shanghai}, \country{China}}}

\affil[3]{\orgdiv{Centre for Audio, Acoustics and Vibration}, \orgname{University of Technology Sydney}, \orgaddress{\city{Ultimo}, \postcode{NSW 2007}, \state{NSW}, \country{Australia}}}

\affil[4]{\orgdiv{Centre for Vision, Speech and Signal Processing}, \orgname{University of Surrey}, \orgaddress{\city{University of Surrey}, \postcode{GU2 7XH}, \country{U.K}}}


\abstract{
Unsupervised anomalous sound detection (ASD) aims to detect unknown anomalous sounds of devices when only normal sound data is available. The autoencoder (AE) and self-supervised learning based methods are two mainstream methods. However, the AE-based methods could be limited as the feature learned from normal sounds can also fit with anomalous sounds, reducing the ability of the model in detecting anomalies from sound. The self-supervised methods are not always stable and perform differently, even for machines of the same type. In addition, the anomalous sound may be short-lived, making it even harder to distinguish from normal sound. This paper proposes an ID constrained Transformer-based autoencoder (IDC-TransAE) architecture with weighted anomaly score computation for unsupervised ASD. Machine ID is employed to constrain the latent space of the Transformer-based autoencoder (TransAE) by introducing a simple ID classifier to learn the difference in the distribution for the same machine type and enhance the ability of the model in distinguishing anomalous sound. Moreover, weighted anomaly score computation is introduced to highlight the anomaly scores of anomalous events that only appear for a short time. Experiments performed on DCASE 2020 Challenge Task2 development dataset demonstrate the effectiveness and superiority of our proposed method.
}

\keywords{Anomalous sound detection, autoencoder, ID classifier, weighted anomaly score computation}

\maketitle

\section{Introduction}\label{sec1}

Anomalous sound detection (ASD) aims to detect anomalies from acoustic signals. Since anomalous sounds can indicate system error or malicious activities, ASD has received much attention \cite{chandola2009anomaly,koizumi2018unsupervised, chalapathy2019deep, nunes2021anomalous, guan2023time}, which has been widely used in various  applications, such as road surveillance \cite{foggia2015audio, li2018anomalous}, animal disease detection \cite{chung2013automatic}, and industrial equipment predictive maintenance \cite{henze2019audioforesight}. Recently, ASD has also been used to monitor the abnormality of industrial machinery equipment, such as anomaly detection for surface-mounted device machine \cite{oh2018residual, park2018fast}, and the Detection and Classification of Acoustic Scenes and Events (DCASE) challenge Task2 from 2020 to 2023 \cite{Koizumi_DCASE2020, Kawaguchi_DCASE2021, Dohi_DCASE2022, Dohi_arXiv2023_01}, to reduce the loss caused by machine damage and the cost of manual inspection. 

Supervised learning based methods usually train a binary classifier to detect the anomaly \cite{li2018anomalous, zabihi2016heart}. However, it is hard to collect enough anomalous data for supervised learning, as actual anomalous sounds rarely occur in real scenarios. In addition, the high diversity of the anomalies can reduce the robustness of supervised methods. Therefore, unsupervised methods are often employed to detect unknown anomalous sounds without using anomalous sound samples.

In unsupervised ASD, a method is to employ the autoencoder (AE) to learn the distributions of sound signals and perform anomaly detection. Conventional AE-based approaches adopt autoencoder to reconstruct multiple frames of spectrogram to learn the distribution of normal sounds, and then the reconstruction error is used to obtain the anomaly score for anomaly detection \cite{oh2018residual, Koizumi_DCASE2020, tagawa2015structured, marchi2015novel, marchi2015non}. However, the conventional AE-based methods do not work well for non-stationary ASD \cite{suefusa2020anomalous}, as non-stationary normal sounds (e.g., sound signals of valves) can easily have larger reconstruction errors than abnormal sounds, thus deteriorating the detection performance. In \cite{suefusa2020anomalous}, an interpolation deep neural network (IDNN) method is proposed, which masks the center frame of the input, and only uses the reconstruction error of the masked center frame to improve non-stationary sound reconstruction, without considering the edge frames. While the method in \cite{wichernanomalous} adopts a similar strategy as IDNN, and applies the local area mask on the input and employs attentive neural process (ANP) \cite{kim2019attentive} for the reconstruction of the masked input.

Instead of reconstructing spectrogram feature, the method in \cite{van2021unsupervised} mixes multiple features as the input, and adopts a fully connected U-Net for the mixed feature reconstruction. To utilize the intra-frame statistics of sound signal, a novel group masked autoencoder for distribution estimation (Group MADE) is proposed for unsupervised ASD \cite{giri2020group, Giri2020}, which estimates the density of an audio time series and achieves better performance. However, the distributions of normal audio clips from different machines are different even for the same sound class. This difference can be even greater than that between normal and anomalous sound, which makes it harder to distinguish normal and anomalous sounds for these purely AE-based methods, as the learned feature from these normal sounds may also fit with the anomalous sounds \cite{Zavrtanik_2021_ICCV}. 

Machine identity (ID) has been used as the additional condition for encoding in the latent feature space of AE, in order to allow the decoder to provide different reconstructions for each machine \cite{2020arXiv200705314K, Kuroyanagi2021}. However, the encoder is unable to learn the difference in distributions for different machines, and as a result, the anomalous sound may be well reconstructed. For this reason, it could still be difficult to distinguish normal and anomalous sound. In addition, the above mentioned AE-based methods often use averaged anomaly score for detection, which does not take into account the short-lived condition in anomalous sound, resulting in low anomaly scores for anomalous events that appear only for a short time, which makes it even more challenging for the AE-based methods.

Therefore, instead of reconstructing normal sounds to learn the feature representation, the self-supervised methods are presented to learn the feature representation by utilizing the difference in distributions among different machines \cite{giri2020self, Wilkinghoff2021, Venkatesh2022, liu2022anomalous, guan2023anomalous, hejing2023interspeech, GuanHEU2022, WeiHEU2022}. The study in \cite{giri2020self} uses machine type and machine ID in addition to the machine condition (normal/abnormal) as training labels for self-supervised classification. The flow-based self-supervised method \cite{dohi2021flow} adopts normalizing flow (NF) \cite{tabak2013family,dinh2014nice} models, such as generative flow (Glow) \cite{kingma2018glow} and masked autoregressive flow (MAF) \cite{papamakarios2017masked}, to obtain the likelihood estimation for anomaly detection. In this method, an auxiliary task is introduced to distinguish the sound data of that machine ID (i.e., target data) from the sound data of other machine IDs with the same machine type (i.e., outlier data). Moreover, although the self-supervised learning based methods can achieve better performance than the AE-based methods, they are not always stable and could perform differently even for the machines of the same type.

In this paper, we present an ID constrained Transformer-based autoencoder (IDC-TransAE) architecture with weighted anomaly score computation for unsupervised ASD. Our method includes two stages, namely, spectrogram reconstruction and anomaly detection.  First, an IDC-TransAE is introduced to reconstruct the spectrogram of normal sounds, where Transformer \cite{vaswani2017attention} is employed to build the AE architecture, and a simple ID classifier is incorporated into the AE. Specifically, the Transformer captures the time-dependent information of the sound signal, and the classifier utilizes machine ID to constrain the latent space of AE, so that our proposed IDC-TransAE can learn different distributions of normal machines, even with the same type.  In the proposed IDC-TransAE architecture, instead of using the positional encoding (PE) for Transformer to provide additional temporal information, a linear phase embedding (LPE) method is proposed to represent the temporal information of sound signal by using its phase information, which can further enhance the classification performance of the proposed IDC-TransAE. In addition, the center frame prediction (CFP) is also employed in our IDC-TransAE to improve the ASD ability for non-stationary signals (e.g., Valve). Then, the reconstruction error from the trained IDC-TransAE can be used to calculate the anomaly score to detect the anomaly. Here, we introduce a weighted anomaly score computation method via global weighted ranking pooling (GWRP) \cite{kolesnikov2016seed}, which can highlight the anomaly scores for the anomalous events that only appear for a short time. Finally, we obtain the final anomaly score with the combination of the classification anomaly score and weighted reconstruction anomaly score, to obtain more stable and consistent detection performance. 

In summary, the innovations and contributions of this paper for unsupervised anomalous sound detection can be summarized as follows:
\begin{enumerate}
    \item We analyze the generalization problem of AE for ASD and point out the main reason for this problem, and propose a solution, i.e., IDC-TransAE, to mitigate the generalization of AE  and improve the detection performance. To the best of our knowledge, this is the first work to clearly point out the main reason for the generalization problem of AE for ASD.
    \item We propose an ID constraint (IDC) classifier to learn different audio feature distributions from the same machine type, which can enhance the distinguishing ability for anomaly detection.
    \item We design a linear phase embedding (LPE) to replace the traditional positional encoding (PE) to preserve the own temporal information of machine sounds by the phase of sounds. 
    \item In the anomaly score calculation, we introduce the global weighted ranking pooling (GWRP) to highlight the anomaly score of sounds with short-time non-stationary anomalies, which obtains a more stable and consistent detection performance.
    \item Experimental results verify that the proposed IDC-TransAE method can mitigate the generalization problem of AE for ASD. Ablation studies and visualizations further verify the effectiveness of the design of ID constraint, LPE and GWRP for ASD. Our study employs the DCASE 2020 Challenge Task2 dataset to address AE's generalization problem in ASD, excluding DCASE 2022 and 2023 datasets tailored for domain-shift and first-shot scenarios beyond our paper's scope.
\end{enumerate}

\section{Preliminary}\label{sec2}

\begin{figure}[htbp]
    \centering
    \centerline{\includegraphics[width=0.85 \textwidth]{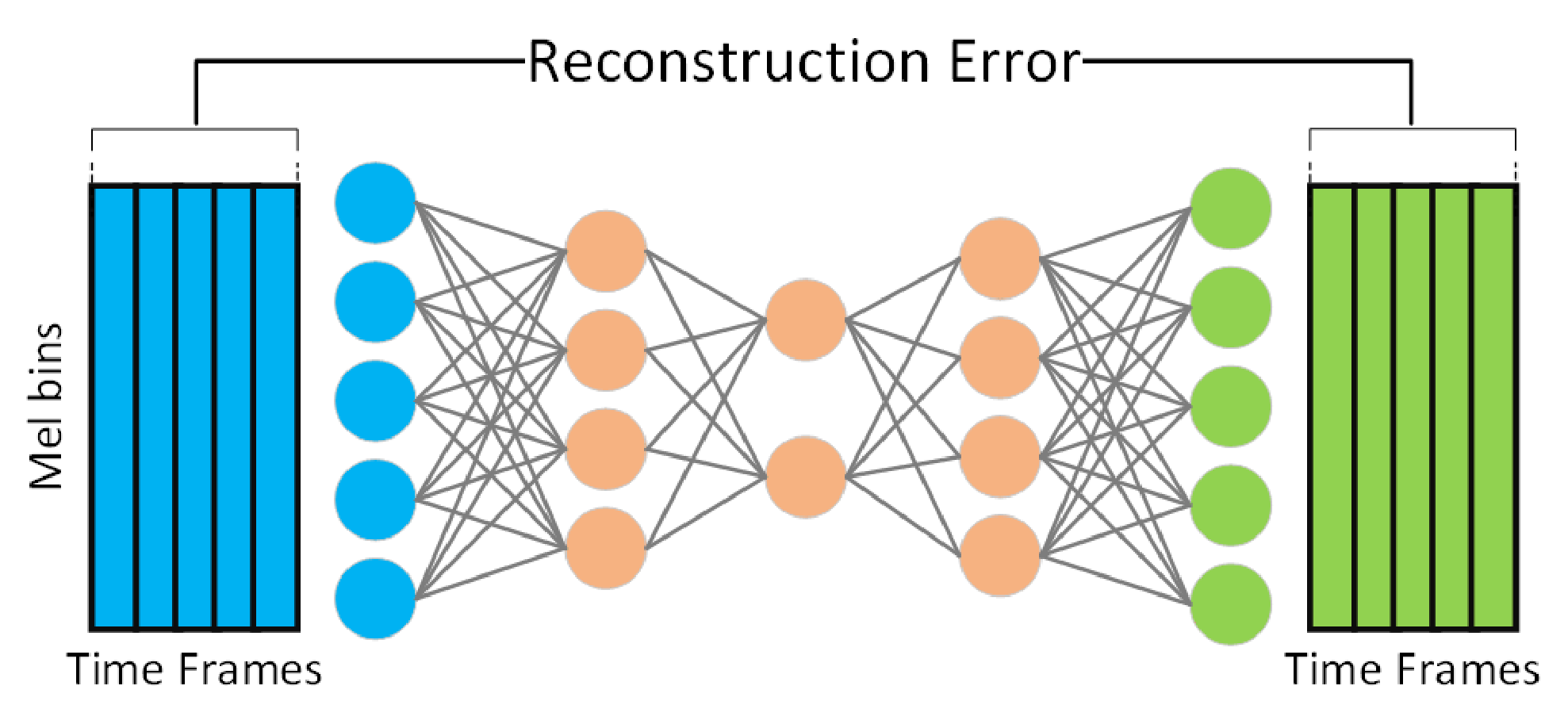}}
	\caption{Typical architecture of AE for unsupervised ASD uses the reconstruction error between the input and output as the anomaly score.}
 \label{fig:1}
\end{figure}

The AE-based methods are widely used for unsupervised ASD \cite{tagawa2015structured, marchi2015novel, oh2018residual, Koizumi_DCASE2020} An AE model is trained with normal sounds to learn their feature distribution. It implicitly assumes that it can reconstruct normal sounds better than anomalous sounds, so that anomalous sounds often have larger reconstruction errors than normal sound. The reconstruction error is then used for deriving the anomaly scores for anomaly detection. Figure ~\ref{fig:1} shows the AE architecture for unsupervised ASD.

Regarding model training, multiple frames of a spectrogram are usually used as the input, and the same number of frames are generated as the output. Suppose $\bm{X} \in \mathbb{R}^{N \times M}$ is the log-Mel spectrogram of the sound signal, where $N$ is the number of frames and $M$ is the feature dimension of each frame of $\bm X$. The loss for the AE model training is
\begin{equation}
\label{eq:1}
    L_{\mathrm{AE}} = \left \| \bm X - D(E(\bm X)) \right \|_{2}^{2},
\end{equation}
where $E(\cdot )$  and $D(\cdot)$ are the encoder and the decoder of AE, respectively.

Then, the trained AE model can be used to detect the anomaly. $\bm{Y}$ is the test audio clip, split into $I$ segments, $\{\bm{Y}_i\}_{i=1}^{I}$. Here $\bm{Y}_i \in \mathbb{R}^{N \times M}$ is the $i$-th segment and also the $i$-th input of the model. The reconstruction error $e_i$ for $\bm{Y}_i$ is
\begin{equation}
\label{eq:2}
e_i = \frac{1}{NM} \left \|( \bm{Y}_{i}-\overline{\bm{Y}}_{i}) \right \|_{F}^2,
\end{equation}
where $\overline{\bm{Y}}_i=D(E(\bm{Y}_i))$ is the corresponding output frames, and $\left \| \cdot \right \|_{F}$ denotes Frobenius norm. It results in a reconstruction error sequence $\bm{e}= \{e_i\}_{i=1}^{I}$ for $\bm{Y}$, and the mean reconstruction error of $\bm{e}$ can be used as the anomaly score
\begin{equation}
\label{eq:3}
 \mathcal A(\bm{e})_{mean} = \frac{1}{I} \sum_{i=1}^{I} e_{i}.
\end{equation}
Here $\mathcal{A}(\bm e)_{mean}$ represents the anomalous degree of the audio clip. The normal or anomaly of the clip is determined by  $\mathcal{H}(\bm e, \theta)$ \cite{koizumi2017optimizing}:
\begin{equation}
\label{eq:4}
 \mathcal H(\bm e, \theta) = \left\{\begin{matrix}
 0 \ (Normal)\ & \mathcal A(\bm e)_{mean} \leq \theta & \\
 1 \ (Anomaly) & \mathcal A(\bm e)_{mean} > \theta & 
\end{matrix}\right.,
\end{equation}
where $\theta$ is a pre-defined threshold value to determine whether an audio clip is anomalous.
\begin{figure}[htbp]
    \centering
    \centerline{\includegraphics[width=0.9\textwidth]{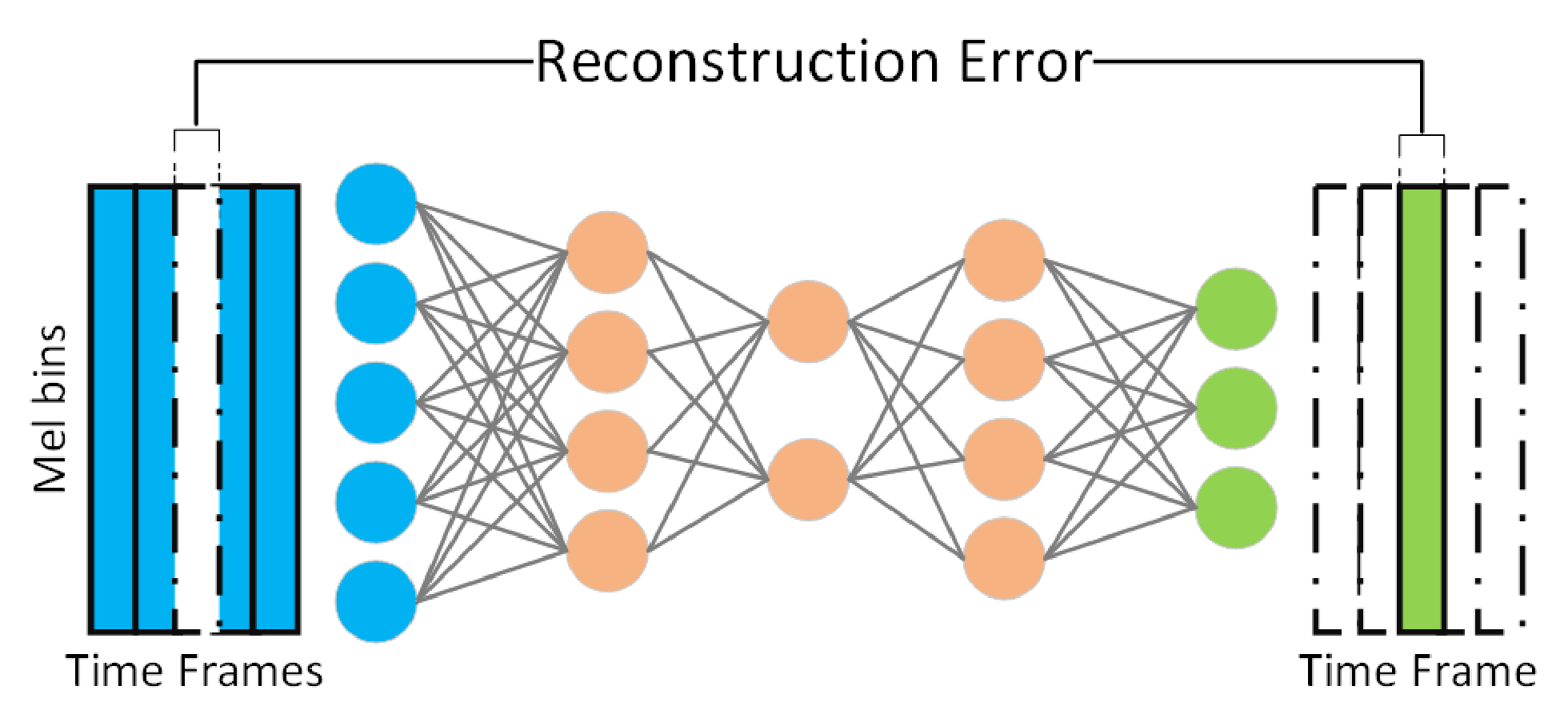}}
	\caption{The architecture of IDNN uses the reconstruction error of center frame as the anomaly score.}
	\label{fig:2}
\end{figure}

However, for normal non-stationary sounds, the AE-based methods tend to give large reconstruction errors for both normal and abnormal sounds, this is because the edge frames of non-stationary sound are hard to reconstruct. In \cite{suefusa2020anomalous}, IDNN is proposed  for non-stationary sound ASD, which removes the center frame of the multiple frames as the input, and predicts the removed frame as the output, as shown in Figure~\ref{fig:2}. The input multiple frames of IDNN can be expressed as $\bm X=[\bm x_1,\cdots,\bm x_{\frac{N+1}{2}-1},\bm x_{\frac{N+1}{2}+1},\cdots,\bm x_N]^{T}$, and $T$ denotes transposition. The loss function of IDNN is formulated as
\begin{equation}
\label{eq:5}
 L_{\mathrm{IDNN}} = \left \| \bm x_{\frac{N+1}{2}} - D(E(\bm X)) \right \|_{2}^{2},
\end{equation}
where $\bm x_{\frac{N+1}{2}}$ is the removed center frame of original input frames. Unlike conventional AE-based methods, the reconstruction error $e_i$ of the $i$-th input is only calculated by the center frame. 

However, the training procedure does not involve the anomalous sound, as a result, the AE-based methods could be limited in the scenario where the learned feature also fits with the anomalous sound \cite{koizumi2018unsupervised}. In this case, the anomalous sound could be well reconstructed with a smaller reconstruction error than that of the normal sounds of different machines, even of the same type. For example, the anomalous sounds from one machine may be similar to the normal sounds of another machine, due to different usage of different machines. In this case, the AE trained with these different machines of the same machine type can reconstruct the anomalous sounds well, and thus it may not be able to detect these anomalous sounds.

In addition, for anomalous events that only appear for a short time in audio clips, the anomaly score calculated by mean reconstruction error is often too small, making it difficult to detect the anomaly.

\section{Proposed Method}
\label{sec:3}
This section presents our IDC-TransAE with weighted anomaly score computation for unsupervised ASD. We introduce IDC-TransAE to reconstruct the spectrogram of normal sounds to learn their distributions, and apply GWRP for weighted anomaly score computation to perform anomaly detection. 
\subsection{ID Constraint Transformer Autoencoder}
\label{sec:3_1}
\begin{figure}[ht]	
	\centering
        \includegraphics[width=1.0\textwidth]{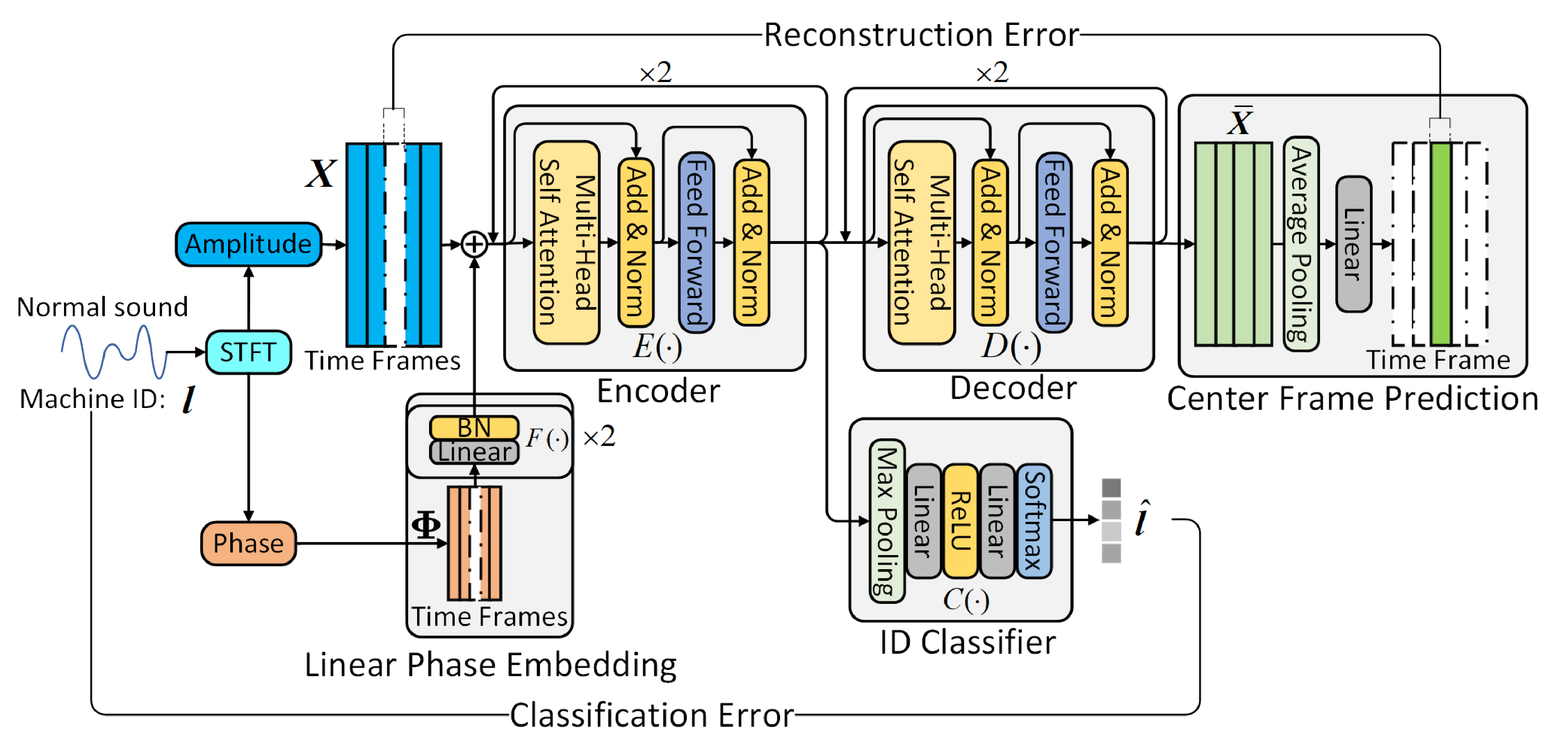}
	\caption{The architecture of the proposed IDC-TransAE for normal sound reconstruction. $\bm X$ and $\bm \Phi$ are the inputs to the model, which are obtained from the sound signal by removing the center frame. The final predicted center frame is obtained by average pooling of the output of the decoder $\overline{\bm{X}}$ in frames and a linear layer, and $\hat{\bm{l}}$ is the predicted machine ID probability of sound signal, which is obtained by max pooling of output of encoder $\bm z$ in frames and two linear layers with softmax. IDC-TransAE is optimized by the combination of reconstruction error and classification error.}
\label{fig:3}
\end{figure}

We utilize Transformer to exploit temporal information for better reconstruction of normal sounds, where only the encoder layer of Transformer is employed to build the encoder and decoder of our IDC-TransAE architecture. In addition, machine ID is adopted to constrain the latent space of the AE by introducing a simple ID classifier to learn different representations for different normal sounds. The framework of the proposed IDC-TransAE is illustrated in Figure~\ref{fig:3}. 

\subsubsection{Center Frame Prediction} 
For better reconstruction of the spectrogram of normal non-stationary sound, following IDNN, we introduce a center frame prediction (CFP) method by removing the center frame of input frames and predicting the removed frame. After removing center frame $\bm x_{\frac{N+1}{2}}$, the input frames can be expressed as 
\begin{equation}
\label{eq:cfp}
\bm X=[\bm x_1,\cdots,\bm x_{\frac{N+1}{2}-1},\bm x_{\frac{N+1}{2}+1},\cdots,\bm x_N]^{T},
\end{equation}
where ${T}$ denotes the matrix transposition operation. Unlike IDNN, the predicted center frame obtained by CFP is the average pooling of decoder output in frames and then processed by a linear layer, as shown in Figure~\ref{fig:3}. 

\subsubsection{Linear Phase Embedding} 
To represent the appropriate positional relationship of the sound signal, we propose a linear phase embedding (LPE) method for IDC-TransAE, to replace positional encoding (PE), often used in Transformer to provide additional position information via sinusoid function \cite{vaswani2017attention}, which is, however, not strongly correlated with the sound signal. In contrast, LPE in the proposed method preserves the signal's temporal information by linearly embedding the phase angles of the signal to the same dimensions with the input $\bm X$. The phase angle is obtained via the short-time Fourier transform (STFT).

Assuming the phase angles corresponding to $\bm X$ are denoted as $\bm \Phi=[\bm \phi_1,\cdots,\\ \bm \phi_{\frac{N+1}{2}-1},\bm \phi_{\frac{N+1}{2}+1},\cdots,\bm \phi_N]^{T}$, with center frame removed, and $F(\cdot)$ is the linear embedding function, including two linear layers with batch normalization.  The output $\overline{\bm{X}}$  of decoder $D(\cdot)$ can be obtained as 
\begin{equation}
\label{eq:6}
\overline{\bm{X}} =  D(E(\bm X + F(\bm \Phi))),
\end{equation}
where $\overline{\bm{X}} = [\overline{\bm{x}}_{1},\cdots,\overline{\bm{x}}_{n},\cdots, \overline{\bm{x}}_{{N-1}}]^{T}$. Then, the average pooling of $\overline{\bm{X}}$ is used to predict the center frame, and the reconstruction loss for the center frame is 
\begin{equation}
\label{eq:7}
 L_{r} = \left \| \bm{x}_{\frac{N+1}{2}} - (W_{o}{\frac{1}{N-1}}\sum_{n=1}^{N-1}{\overline{\bm{x}}_{n}}+b_{o}) \right \|_2^2,
\end{equation}
where $W_{o}$ and $b_{o}$ are the learnable parameters of the last output linear layer. 
The LPE module helps preserve the temporal information of the signal to enhance the ability of the model for anomalous sound detection. 

\subsubsection{ID Classifier}

We observed that the performance of the trained AE model on different machines with the same type could be quite different. The potential cause is the difference in distributions of normal machine sound when individual machines have different usages. However, the trained model only learns how to reconstruct the general distribution of different normal machines sounds.

To enable the model to learn different representations for different machine sounds even with the same type, we introduce an ID classifier $C(\cdot)$ with machine ID information to constrain the latent feature $\bm z$ of AE. The structure of the ID classifier $C(\cdot)$ is given in Figure \ref{fig:3}, which consists of a max pooling layer, two linear layers with a ReLU \cite{glorot2011deep} function and a softmax activation function.

Here, the latent feature $\bm z$ is the output of the encoder of AE, which is the input of the classifier, defined as $\bm z=E(\bm X+F(\bm \Phi))$. The output of the ID classifier $\hat{\bm {l}} = C(\bm z) \in \mathbb{R}^K$ is the probability indicating normal/anomalous sound corresponding to the machine ID, and $K$ is the number of machines with the same type.
Then, the classification error of $C(\cdot)$ can be obtained via a cross-entropy loss function~\cite{murphy2012machine}
\begin{equation}
\label{eq:8}
L_{c} = CrossEntropy(\bm l, \hat{{\bm{l}}}),
\end{equation}
where $\bm l \in \mathbb{R}^K$ is the one-hot vector of machine ID label of the sound signal.

Therefore, the proposed IDC-TransAE can be jointly trained by minimizing the center frame reconstruction error and the machine ID classification error with the joint loss function
\begin{equation}
\label{eq:9}
L_{total} = (1-\alpha)L_{r} + \alpha L_{c},
\end{equation}
where $\alpha \in [0,1) $ is a hyper-parameter. The magnitude of $\alpha$ denotes the extent to which the machine ID classifier restricts $\bm z$. By jointly training the AE with the ID classifier, we can improve anomaly detection performance.

\subsection{Weighted Anomaly Score Computation}
\label{sec:3_2}
For anomaly detection, the formula $\mathcal{A}(\bm{e})_{mean}$ in Equation \eqref{eq:3} usually underestimates the anomaly scores of anomalous audio clips when the anomalous events only appear for a short time. One solution is to use the maximal reconstruction error as the anomaly score i.e., max anomaly score $\mathcal{A}(\bm{e})_{max}=max(\bm{e})$, to highlight the anomalies of these audio clips. However, it is not robust to use the maximum value of $\bm{e}$ as the anomaly score of the whole audio clip, as it may overestimate the anomaly scores of some normal audio clips.

To improve the reliability of the calculated anomaly score, we employ the global weighted rank pooling (GWRP) method to obtain weighted anomaly score, where GWRP is a generalization of max and mean, which can highlight the anomaly score by setting different weights to reconstruction error sequence $\bm{e}$. For example, let $\hat{\bm{e}}=\{\hat{e}_1, ..., \hat{e}_I\}$ be sorted by descending order of $\bm{e}$, the GWRP anomaly score can be calculated as
\begin{equation}
\label{eq:10}
\mathcal A(\hat{\bm{e}})_{gwrp} = \frac{1}{Z(r)} \sum_{i=1}^{I} r^{i-1} \hat e_i,
\end{equation}
where $0 \leq r \leq 1$ is a hyper-parameter and $Z(r)=\sum_{i=1}^I r^{i-1}$ is a normalization term. When $r=0$, $\mathcal A(\hat{\bm{e}})_{gwrp}$ degenerates to $\mathcal A(\bm{e})_{max}$, and when $r=1$,  $\mathcal A(\hat{\bm{e}})_{gwrp}$ becomes $\mathcal A(\bm{e})_{mean}$. It intends to assign larger weights to anomalous audio clips and lower weights to normal audio clips, to generate high anomaly scores for the anomalous events of short duration. In addition, the classification error is combined with the reconstruction error to calculate the anomaly score, to allow the anomaly score to increase if the ID classifier misclassifies the machine ID. Finally, the weighted anomaly score can be calculated as
\begin{equation}
\label{eq:14}
\mathcal A(\hat{\bm{e}}, \bm l, \hat{\bm{l}}) = \mathcal (1 - \beta)\mathcal A(\hat{\bm{e}})_{gwrp} + \beta  L_{c},
\end{equation}
where $\beta \in [0, 1]$ is a parameter weighting the impact of a false prediction by the ID classifier on the anomaly score. For clarity, the proposed IDC-TransAE with weighted anomaly score computation is denoted as IDC-TransAE-W in the following section.

\section{Experiments and Results}
\label{sec:4}
\subsection{Experimental Setup}
\label{sec:4_1}
\subsubsection{Dataset} 
We evaluate our method on the DCASE 2020 Challenge Task2 \cite{Koizumi_DCASE2020} dataset, which comprises parts of  MIMII  \cite{Purohit_DCASE2019_01} and ToyADMOS dataset \cite{Koizumi_WASPAA2019_01} including the normal/anomalous operating sounds of six types of real/toy machines. The MIMII dataset includes four types of machines (i.e., Fan, Pump, Slider and Valve), with four different machines for each machine type. The ToyADMOS dataset consists of two types of machines (i.e., ToyCar and ToyConveyor), with four and three different machines for each type, respectively. Each recording is a single-channel audio of 10-sec long with a 16kHz sampling rate that includes both a target machine's operating sound and environmental noise. Following \cite{Koizumi_DCASE2020}, the training set only includes normal sounds, with around 6000 items for each machine type, and the test set consists of both normal and anomalous sounds, including about 500 to 1000 items for normal and anomaly in each machine type.
\subsubsection{Performance Metrics} 
Following \cite{Koizumi_DCASE2020, suefusa2020anomalous, perez2020anomalous, giri2020self, dohi2021flow}, we employ area under the receiver operating characteristic (ROC) curve (AUC) and the partial-AUC (pAUC) as the performance metrics, where the pAUC is calculated as the AUC over a low false-positive-rate (FPR) range $[0, p]$ and $p=0.1$ as in \cite{Koizumi_DCASE2020}.  Higher AUC indicates better model performance. pAUC reflects the reliability of the ASD system based on practical requirements. It is important to increase pAUC to avoid the ASD system predicting false alerts frequently \cite{Koizumi_DCASE2020}. In addition, the minimum AUC (mAUC) is adopted to represent the worst detection performance achieved among individual machines of same machine type, following \cite{dohi2021flow}.
\subsubsection{Implementation Details} 
\begin{table*}[h]
\centering
\footnotesize
    \vspace{-3mm}
	\caption{Implementation details for all machine types.}
	~\\
    \begin{tabular}{c|cccccc}
    \hline
               & \multicolumn{1}{c|}{Fan}  & \multicolumn{1}{c|}{Pump} & \multicolumn{1}{c|}{Slider} & \multicolumn{1}{c|}{Valve} & \multicolumn{1}{c|}{ToyCar} & ToyConveyor \\ \hline
    n\_FFT    & \multicolumn{6}{c}{1024}                                                                                                                                     \\ 
    n\_Mels   & \multicolumn{6}{c}{128}                                                                                                                                      \\ 
    hop length & \multicolumn{6}{c}{512}                                                                                                                                      \\ 
    frames     & \multicolumn{6}{c}{5}                                                                                                                                        \\ \hline
    $\alpha$   & \multicolumn{6}{c}{0.3}                                                                                                                                      \\ 
    $r$        & \multicolumn{1}{c}{1.00}  & \multicolumn{1}{c}{1.00}  & \multicolumn{1}{c}{0.96}   & \multicolumn{1}{c}{0.92}  & \multicolumn{1}{c}{1.00}   & 1.00         \\ 
    $\beta$    & \multicolumn{1}{c}{0.84} & \multicolumn{1}{c}{0.82} & \multicolumn{1}{c}{0.80}   & \multicolumn{1}{c}{0.72}  & \multicolumn{1}{c}{0.62}   & 0.98        \\ \hline
    \end{tabular}
\label{tab:1}
\end{table*}
The implementation details of IDC-TransAE can be seen in Table \ref{tab:1}. 
We use the log-Mel spectrogram and phase angle of the sound signal as the input of our IDC-TransAE. The frame size is 1024 with an overlapping 50\%, i.e., the number of FFT bins (n\_FFT) is 1024, and the hop length is 512. The number of Mel filter banks (n\_Mels) is set as 128. The number of frames (i.e., $N$) is 5. The dimension of phase angles is 513, which is embedded to a 128-dimensional vector by the linear function $F(\cdot)$. Here,  $F(\cdot)$ consists of two linear layers with batch normalization. The encoder and decoder of IDC-TransAE include two layers, respectively. The classifier includes a max pooling layer, two linear layers with a ReLU  and a softmax activation function. The hyper-parameter $\alpha$ of the joint loss function is empirically set as 0.3. 

Adam optimizer \cite{kingma2014adam} is used to optimize our model with a learning rate of 0.0001. For each machine type, our model is trained 300 epochs, and the batch size is set as 2000. In the joint training stage, we found that the classification loss converges much faster than the reconstruction loss, 
so we adopt a training strategy to avoid the overfitting of the classifier, by training the classifier every 10 epochs (i.e., using $L_{total}$ loss) and the remaining epochs for autoencoder (i.e., using $L_{r}$ loss). In weighted anomaly score computation, $r$ and $\beta$ are empirically selected, and the values of $r$ and $\beta$ are provided in Table \ref{tab:1}. 

\subsection{Experimental Results and Performance Analysis}
\subsubsection{Comparison with Other Methods}
\label{sec:4_2}
To demonstrate the performance of our method for unsupervised ASD, we compare our approach with the AE baseline of DCASE 2020 Challenge Task2  \cite{Koizumi_DCASE2020} and mainstream models, including AE-based methods (i.e.,  IDNN \cite{suefusa2020anomalous}, ANP-Boot \cite{perez2020anomalous}, Group MADE \cite{giri2020group} and IDCAE \cite{2020arXiv200705314K}) and self-supervised based methods (i.e.,  MobileNetV2 \cite{giri2020self} and Glow\_Aff \cite{dohi2021flow}), where IDCAE, MobileNetV2 and Glow\_Aff  employ the ID information for anomalous sound detection.

\begin{table*}[htbp]
\footnotesize
    \centering
    \Huge
	\caption{Performance comparison in terms of AUC (\%) and pAUC (\%) for different types of machines.}
	\resizebox{\textwidth}{!}
	{
		\begin{tabular}{ccccccccccccccc}
			\toprule
			\multirow{2}{*}{} &\multicolumn{2}{c}{Fan}&\multicolumn{2}{c}{Pump}&\multicolumn{2}{c}{Slider}&\multicolumn{2}{c}{Valve}&\multicolumn{2}{c}{ToyCar}&\multicolumn{2}{c}{ToyConveyor}&\multicolumn{2}{c}{Average} \\
			\cmidrule(r){2-3} \cmidrule(r){4-5} \cmidrule(r){6-7} \cmidrule(r){8-9} \cmidrule(r){10-11} \cmidrule(r){12-13} \cmidrule(r){14-15} 
			  & {AUC} & {pAUC} & {AUC} & {pAUC} & {AUC} & {pAUC} & {AUC} & {pAUC}& {AUC} & {pAUC} & {AUC} & {pAUC} & {AUC} & {pAUC} \\
			\midrule
			\multicolumn{3}{l}{w/o ID information }\\
			\midrule
			AE baseline \cite{Koizumi_DCASE2020} & 65.91 & 51.93 & 70.20 & 61.69 & 83.42 & 65.72 & 67.78 & 51.67 & 78.77 & 67.58 & 72.53 & \textbf{60.43} & 73.10 & 59.84 \\
			IDNN \cite{suefusa2020anomalous} & 65.94 & 52.48 & 74.26 & 62.20 & 84.34 & 65.48 & 83.70 & 62.02 & 77.42 & 62.64 & 69.36 & 58.58 & 75.67 & 60.57 \\
			ANP-Boot \cite{wichernanomalous} & 64.80 & 53.00 & 65.50 & 59.00 & \textbf{94.90} & 83.10 & 85.20 & 72.00 & 72.90 & 68.10 & 67.10 & 54.20 & 75.07 & 64.90 \\
			Group MADE \cite{giri2020group} & 68.00 & 53.10 & 74.10 & 66.20 & 94.40 & \textbf{83.70} & 95.60 & 85.50 & 79.50 & 68.40 & \textbf{74.70} & 60.30 & 81.05 & 69.53 \\
			\textbf{TransAE-mean} & 73.91 & 54.14 & 77.31 & 68.96 & 91.51 & 74.66 & 96.09 & 84.65 & 80.62 & 72.65 & 74.32 & 59.80 & 82.29 & 69.14 \\
			\textbf{TransAE-W} & \textbf{73.91} & \textbf{54.14} & \textbf{77.31} & \textbf{68.96} & 94.52 & 82.33 & \textbf{99.68} & \textbf{98.31} & \textbf{80.62} & \textbf{72.65} & 74.32 & 59.80 & \textbf{83.39} & \textbf{72.70} \\
			\midrule
			\multicolumn{3}{l}{w/ ID information} \\
			\midrule
			MobileNetV2 \cite{giri2020self}
			& 80.19 & \textbf{74.40}  & 82.53 & 76.50  & 95.27 & 85.22  & 88.65 & 87.98  & 87.66 & 85.92  & 69.71 & 56.43  & 84.34 & 77.74 \\
    		Glow\_Aff \cite{dohi2021flow}
    		& 74.90 & 65.30  & 83.40 & 73.80  & 94.60 & 82.80  & 91.40 & 75.00  & 92.20 & 84.10 
    		& 71.50 & 59.00 
    		& 85.20 & 73.90 \\
    		IDCAE \cite{2020arXiv200705314K}
    		& 77.45 & 70.32  & 77.29 & 70.33  & 80.04 & 68.25  & 78.26 & 55.80  & 78.07 & 74.22 
    		& 70.29 & 59.46 
    		& 76.90 & 66.40 \\
    		\textbf{IDC-TransAE-mean} & 80.44 & 70.21 & 83.41 & 79.24 & 92.17 & 77.10 & 94.04 & 78.94 & 93.17 & 87.43 & 75.69 & 62.96 & 86.49 & 75.98 \\
    		\textbf{IDC-TransAE-W} & \textbf{80.44} & 70.21 & \textbf{83.41} & \textbf{79.24} & \textbf{96.20} & \textbf{86.38} & \textbf{99.60} & \textbf{98.29} & \textbf{93.40} & \textbf{87.43} & \textbf{75.69} & \textbf{62.96} & \textbf{88.12} & \textbf{80.94} \\
			\bottomrule
			\bottomrule
			\end{tabular}
	}
	\label{tab:2}
\end{table*}

Table \ref{tab:2} shows the comparison results in terms of AUC and pAUC. Here,  IDC-TransAE-W and IDC-TransAE-mean represent IDC-TransAE with weighted anomaly score computation and mean anomaly score computation, respectively. In addition, the proposed methods without using ID information are evaluated, i.e., TransAE-W and TransAE-mean.

As shown in Table \ref{tab:2}, the methods with ID information (denoted as w/ ID) give better detection performance than the methods without ID information (denoted as w/o ID), except IDCAE. The proposed IDC-TransAE-W performs the best in terms of average AUC and pAUC. Amongst the methods without using ID information, our TransAE-W also achieves the best overall performance. Especially, both TransAE-W and IDC-TransAE-W can substantially improve the performance on the non-stationary sound signal of Valve (i.e., with 13.66\% and 19.35\% pAUC improvements compared to TransAE-mean and IDC-TransAE-mean, respectively), which demonstrates the effectiveness of the weighted anomaly score computation for anomalous events appearing for a short time. In addition, the significantly improved average pAUC (i.e., 80.94\%) shows that the proposed IDC-TransAE-W is more reliable than other methods.

Note that, $r$ in the weighted anomaly score computation can be adjusted according to the time length of the anomalous event, for example, when $r=1$, $A(\hat{\bm{e}})_{gwrp} = A(\bm{e})_{mean}$. This means it is more applicable than mean anomaly score. The influence of $r$ will be discussed in Section \ref{sec:visual}.

\subsubsection{Detection Stability}
\label{sec:4_2_2}
To demonstrate the effectiveness of our method for more stable detection, another experiment is conducted to show the worst detection performance on individual machines of the same type, where the self-supervised based methods (i.e.,  MobileNetV2 and Glow\_Aff) and the typical AE-based method (i.e., IDNN) are employed for comparison. The results in terms of mAUC are given in Table \ref{tab:3}.  
\begin{table}[htbp]
\centering
\footnotesize
	\caption{Performance comparison in terms mAUC (\%) among the individual machines of the same type.}
    {\begin{tabular}{ccccc}
        \toprule
                    & MobileNetV2\cite{giri2020self}& Glow\_Aff\cite{dohi2021flow}&
                    IDNN\cite{suefusa2020anomalous}
                    & \textbf{IDC-TransAE-W} \\ \midrule
        Fan & 50.40 & 49.60 & \textbf{56.56} & 50.55 \\
        Pump & 52.90 & \textbf{65.70} & 61.86 & 57.27          \\
        Slider & 82.80 & 87.80 & 74.22 & \textbf{88.64} \\
        Valve & 67.90 & 77.70 & 66.83 & \textbf{99.24} \\
        ToyCar & 55.70 & 80.10 & 64.41 & \textbf{81.35} \\
        ToyConveyor & 48.70 & 61.00 & \textbf{62.89} & 62.31 \\ 
        \midrule
        Average & 59.73 & 70.32 & 64.46 & \textbf{73.23} \\
    \bottomrule
    \bottomrule
    \end{tabular}
    }
\label{tab:3}
\end{table}

As can be seen from Table \ref{tab:2} and Table \ref{tab:3},  the self-supervised methods, i.e., MoblieNetV2 and Glow\_Aff, can achieve significant improvements in average AUC and pAUC, as compared to the AE-based method IDNN. However, they perform dramatically different even for the machines of the same type, as observed from Table \ref{tab:3} and Table \ref{tab:2}, e.g., MobileNetV2 has much smaller mAUC than AUC on  Fan, Pump, ToyCar and ToyConveyor. The results demonstrate the instability of the self-supervised methods. 

Especially, the average mAUC (i.e., 59.73\%) of MobileNetV2 is lower than that of IDNN (i.e., 64.46\%), which indicates that the self-supervised classification method (i.e., MobileNetV2) indeed easily fails on some individual machines and lacks performance consistency. In contrast, the AE-based method IDNN can provide a relatively stable detection performance. Although the flow-based self-supervised method (Glow\_Aff) can improve detection stability to some extent compared to the AE-based method, our proposed method can achieve the best average mAUC performance and obtain more stable performance for some machine types, i.e., Slider, Valve, and ToyCar. 

Although Glow\_Aff has a higher mAUC on Pump than our proposed method, the model needs to be trained for each individual machine which could be limited in real-world applications. In contrast, our proposed method only needs to train one model for each machine type.

\subsubsection{Generalization to Anomaly}

To demonstrate the proposed IDC-TransAE can mitigate the generalization of AE for anomalous sound and improve its detection performance, experiments are conducted to compare it with the typical AE-based method (i.e., IDNN).
\begin{figure}[!htbp]
    \vspace{-5mm}
    \centering
    \subfloat[IDNN]{
    \label{fig:4a}
    \includegraphics[width=0.48\textwidth]{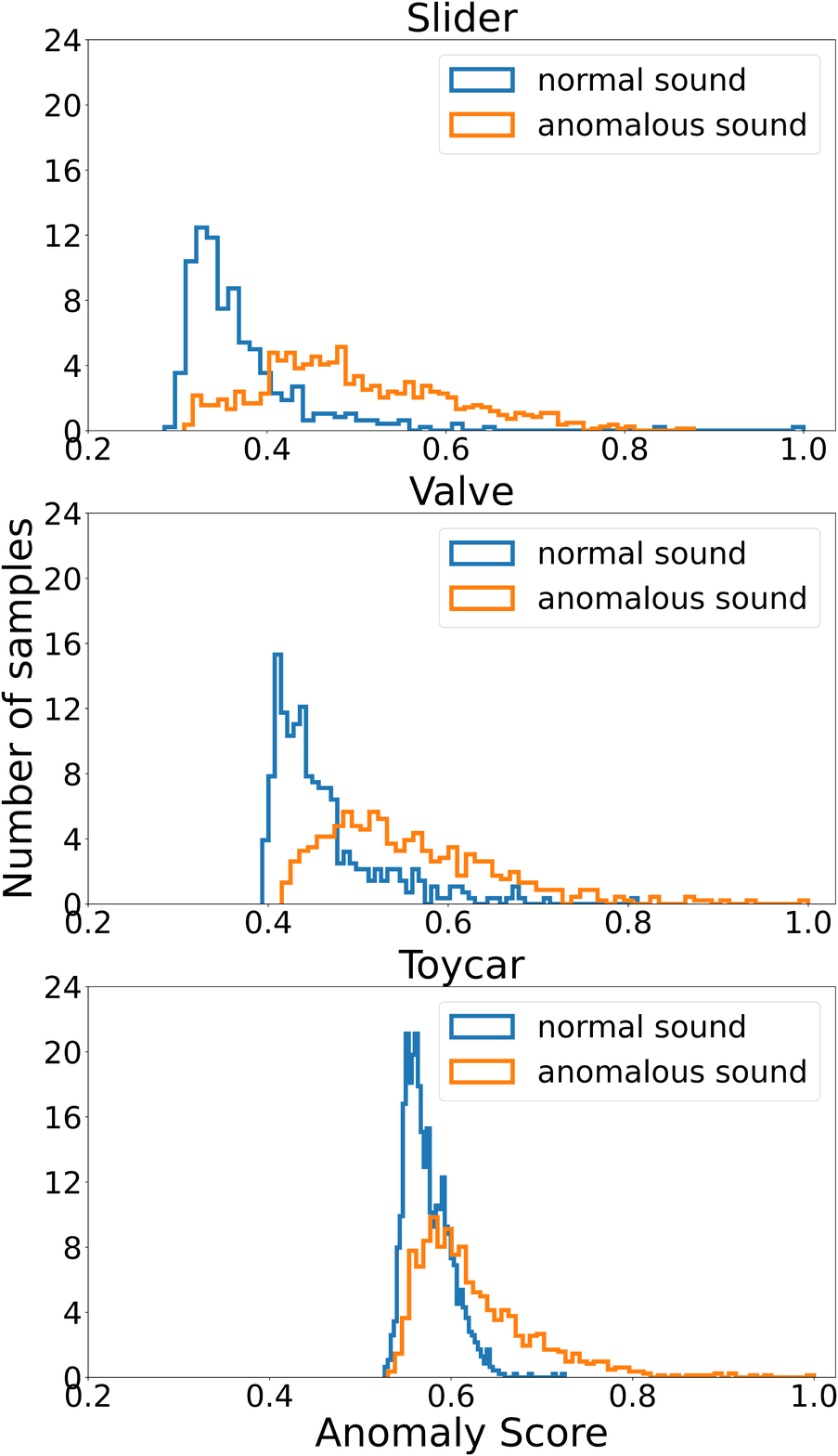}
    }
    \subfloat[IDC-TransAE-mean]{
    \label{fig:4b}
    \includegraphics[width=0.48\textwidth]{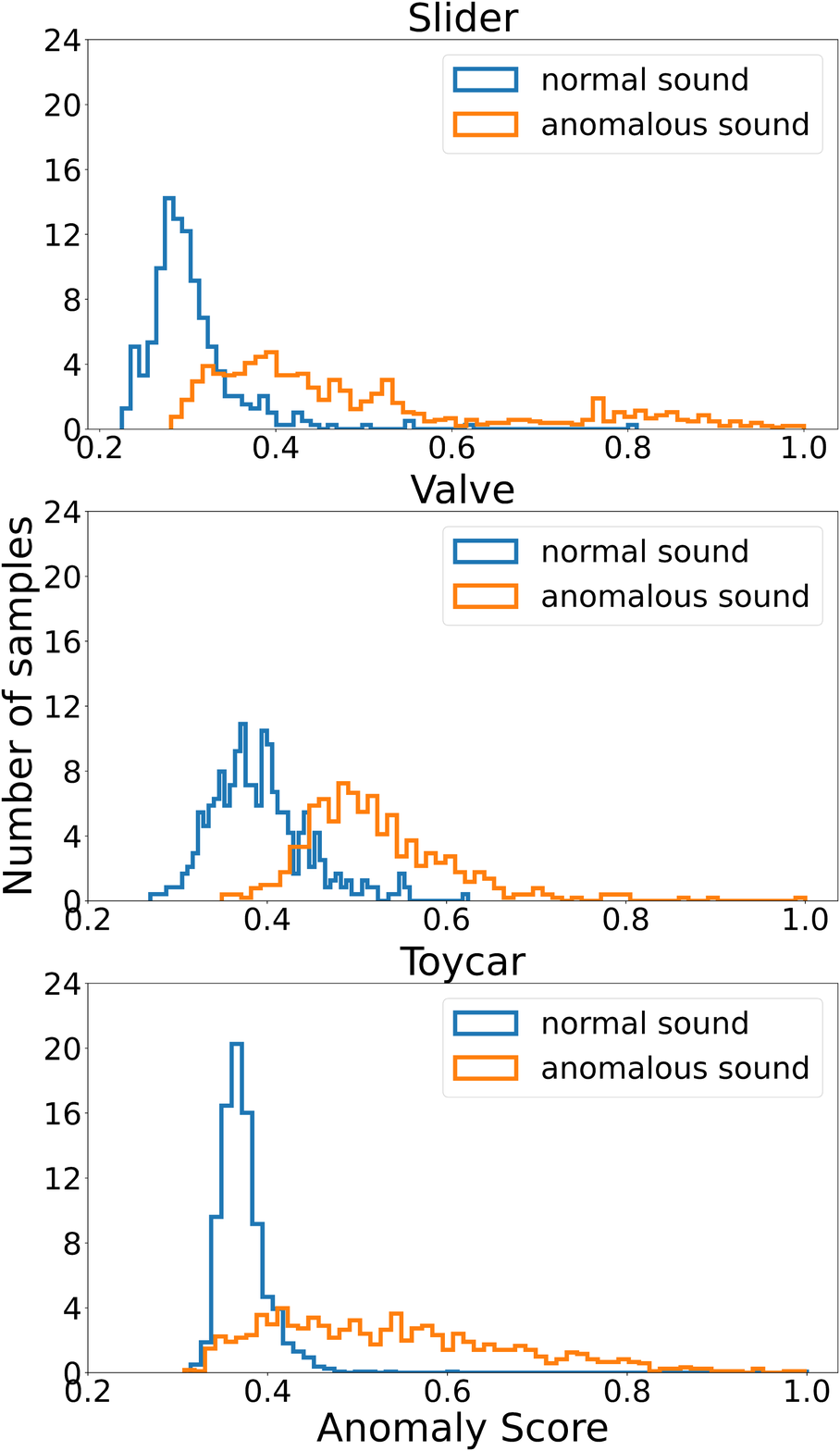}
    }
    \caption{The histograms of anomaly scores distribution on Slider, Valve and ToyCar using IDNN and the proposed IDC-TransAE-mean, where blue and orange indicate the anomaly score distribution of normal and anomalous sound, respectively.}
    \vspace{-5mm}
\label{fig:4}
\end{figure}

First, we show the histograms of anomaly score distribution on Slider, Valve and ToyCar using IDNN and our proposed IDC-TransAE. For a fair comparison, our method (i.e., IDC-TransAE-mean) also adopts mean anomaly score computation as IDNN, and the results are provided in Figure~\ref{fig:4}. Here,  the anomaly score is on the horizontal axis of the histogram, which is normalized to facilitate comparison. The vertical axis represents the number of audio samples corresponding to the anomaly score distribution on the histogram. 

\begin{figure}[htbp]
    \centering
    \subfloat[IDNN for normal sound]{
    \label{fig:5a}
    \includegraphics[width=0.99\textwidth]{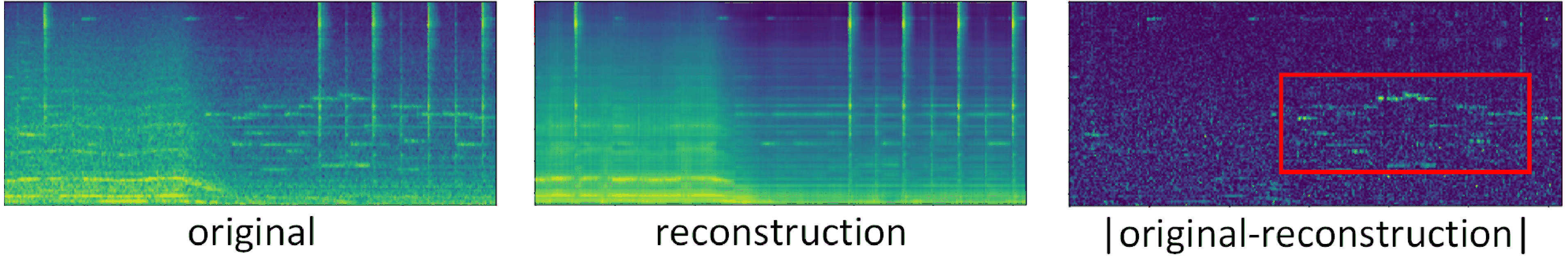}
    }
    \\
    \subfloat[IDC-TransAE for normal sound]{
    \label{fig:5b}
    \includegraphics[width=0.99\textwidth]{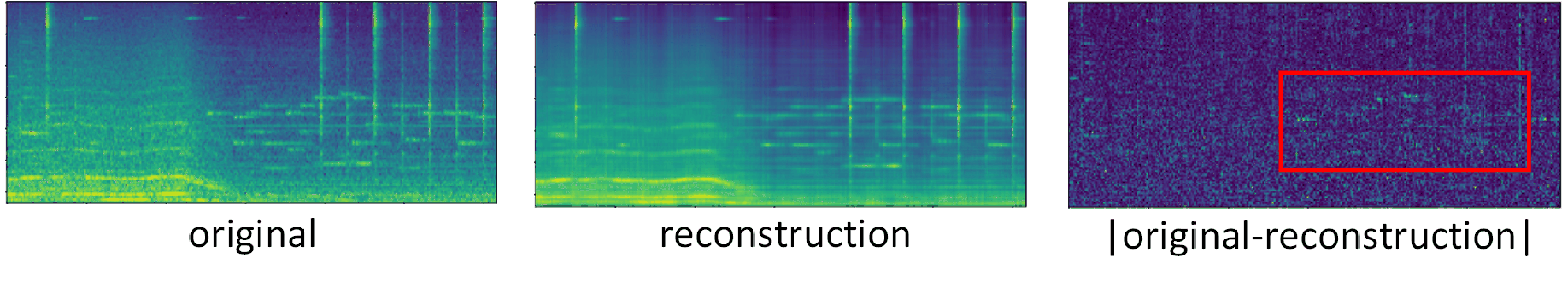}
    }
    \\
    \subfloat[IDNN for anomalous sound]{
    \label{fig:5c}
    \includegraphics[width=0.99\textwidth]{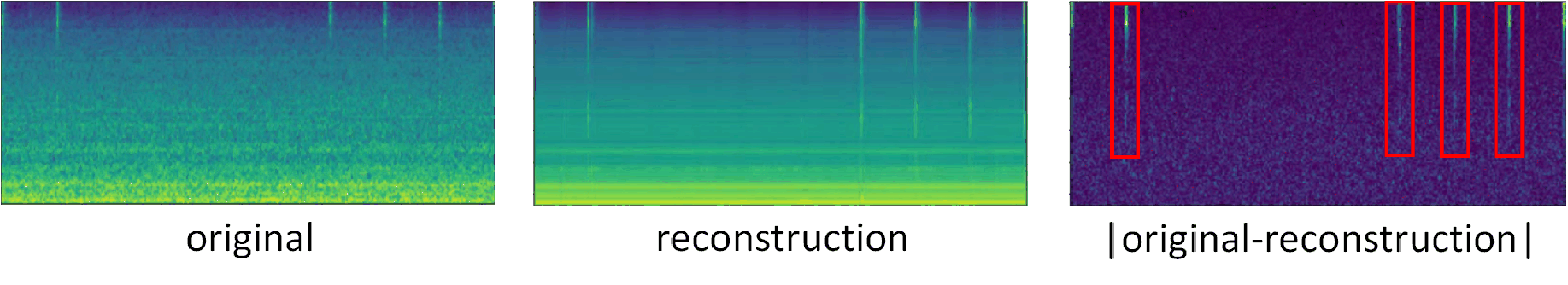}
    }
    \\
    \subfloat[IDC-TransAE for anomalous sound]{
    \label{fig:5d}
    \includegraphics[width=0.99\textwidth]{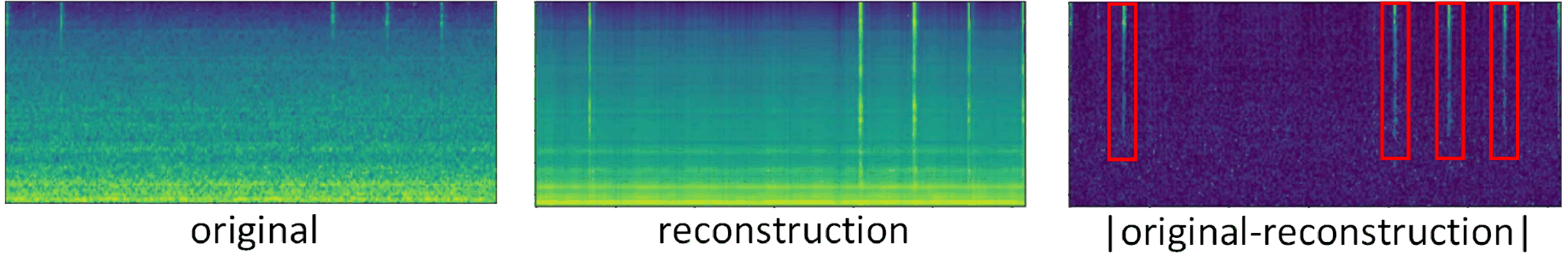}
    }
    \caption{The log-Mel spectrogram reconstruction analysis of IDNN and IDC-TransAE on normal and anomalous Valve's sound, the ``original", ``reconstruction" and ``\textbar original-reconstruction\textbar" represent  original spectrogram, reconstructed spectrogram and the  absolute value of their difference, respectively.} 
    \vspace{-5mm}
\label{fig:5}
\end{figure}
From Figure~\ref{fig:4}, we can see that, for IDNN,  the anomaly score distribution of the anomalous sound tends to be similar to that of the normal sound, especially on ToyCar, as shown in Figure~\ref{fig:4}(a). It shows that most anomalous sound have a small anomaly score similar to normal sound. This indicates that the AE-based method (i.e., IDNN) is able to generalize the representation for anomalous sound, which reduces its ability to distinguish between normal and abnormal sound. In contrast, our proposed method can give higher anomaly scores for the anomalous sound, and provide better detection ability than the AE-based method, as shown in the histograms in Figure~\ref{fig:4}(b), which demonstrates the effectiveness of our proposed IDC-TransAE architecture.

To further demonstrate that our proposed IDC-TransAE can mitigate its generalization for the anomaly, we perform another experiment for non-stationary anomalous sound detection (i.e., sound of Valve) as compared with IDNN, where the log-Mel spectrogram reconstruction of normal and anomalous sound is illustrated in Figure~\ref{fig:5}. From left to right, Figure~\ref{fig:5} shows the original log-Mel spectrograms, the reconstructed log-Mel spectrograms, and the absolute values of their difference.

Comparing the red box areas illustrated in Figure~\ref{fig:5}(a) and Figure~\ref{fig:5}(b), the proposed IDC-TransAE can provide better normal sound reconstruction, as it can achieve  smaller reconstruction error for  normal sound than that of the typical AE-based method (i.e., IDNN). This can be clearly observed in the comparison of the absolute value difference of original log-Mel spectrogram and reconstrcuted log-Mel spectrogram, as the red box indicated areas in Figure~\ref{fig:5}(a) and Figure~\ref{fig:5}(b). Whereas for the anomalous sound reconstruction, our proposed method can give larger reconstruction error than the typical AE-based method, which means that our method has a better ability to highlight the anomalies when reconstructing the anomalous sound. This can be observed from the comparison between the red box areas in Figure~\ref{fig:5}(c) and Figure~\ref{fig:5}(d), where the absolute value difference shown in  Figure~\ref{fig:5}(d) is much more clear than that in Figure~\ref{fig:5}(c). The results further demonstrate that our proposed IDC-TransAE can solve the generalization problem of the AE-based method and has a better ability in anomaly detection.  

Note that the log-Mel spectrogram of the anomalous sound also shows that the anomalies may appear for a short time in the sound, as illustrated in Figure~\ref{fig:5}. In this case, the mean anomaly score computation method will give low anomaly scores for the anomalous events that only appear for a short time.

\subsection{Ablation Studies}
\label{sec:4_3}

To show the effectiveness of different parts of our proposed IDC-TransAE-W,  ablation studies are conducted, where AUC and pAUC are used as performance metric. The results are given in Table \ref{tab:4}. Here, TransAE/PE-W denotes the proposed model without using machine ID constraint (IDC) module and CFP module, and adopts PE instead of LPE, with weighted anomaly score computation for anomaly detection. TransAE/PE/CFP-W denotes the TransAE/PE-W using CFP, and TransAE/LPE/CFP-W denotes replacing PE with LPE in Transformer/PE/CFP-W.
\begin{table*}[htbp]
    \centering
    \caption{Validation of different modules of IDC-TransAE.}
    \label{tab:4}
    \resizebox{\textwidth}{!}{
    \begin{tabular}{ccccccccc}
        \toprule
         & \multicolumn{2}{c}{TransAE/PE-W} & \multicolumn{2}{c}{TransAE/PE/CFP-W} & \multicolumn{2}{c}{TransAE/LPE/CFP-W} & \multicolumn{2}{c}{IDC-TransAE-W} \\
         \cmidrule(r){2-3} \cmidrule(r){4-5} \cmidrule(r){6-7} \cmidrule(r){8-9}
         & AUC & pAUC & AUC & pAUC & AUC & pAUC & AUC & pAUC \\
        \midrule
        Fan & 70.06 & 52.73 & 72.46 & 52.95 & 73.91 & 54.14 & \textbf{80.44} & \textbf{70.21} \\
        Pump & 79.76 & 70.79 & 77.66 & 71.30 & 77.31 & 68.96 & \textbf{83.41} & \textbf{79.24} \\
        Valve & 92.88 & 81.05 & 94.12 & 82.16 & 94.52 & 82.33 & \textbf{96.20} & \textbf{86.38} \\
        Slider & 85.10 & 62.91 & 97.71 & 93.31 & \textbf{99.68} & \textbf{98.31} & 99.60 & 98.29 \\
        ToyCar & 82.95 & 72.47 & 81.27 & 72.69 & 80.62 & 72.65 & \textbf{93.40} & \textbf{87.43} \\
        ToyConveyor & \textbf{76.81} & \textbf{63.88} & 74.35 & 58.51 & 74.32 & 59.80 & 75.69 & 62.96 \\
        \midrule
        Average & 81.26 & 67.31 & 82.93 & 71.82 & 83.39 & 72.70 & \textbf{88.12} & \textbf{80.94}\\
        \bottomrule
    \end{tabular}
    }
\end{table*}

As shown in Table \ref{tab:4}, TransAE/PE/CFP-W can significantly improve the detection performance for the non-stationary sound signal of Valve, with 12.61\% AUC and 30.40\%  pAUC improvements as compared with TransAE/PE-W. To show the effectiveness of LPE, we compare the performance of TransAE/PE/CFP-W and TransAE/LPE/CFP-W. The result shows that TransAE/LPE/CFP-W can improve the detection performance on Fan, Slider, Valve, ToyConveyor, and achieve better average AUC and pAUC performance. It indicates that LPE can better represent the temporal information of the sound signal by using its phase information. By introducing the ID classifier, the proposed IDC-TransAE-W with the IDC module can achieve the best overall detection performance,  giving more than 10\% improvement in pAUC on Fan, Pump, and ToyCar as compared with TransAE/LPE/CFP-W. Besides, we can see that the proposed IDC module contributes the most to the performance improvement in Table \ref{tab:4}, which further verifies the effectiveness of the proposed IDC module to enhance the ability of the model in distinguishing anomalous sound.
\begin{figure}[htbp]
    \centering
    \centerline{\includegraphics[width=0.8 \textwidth]{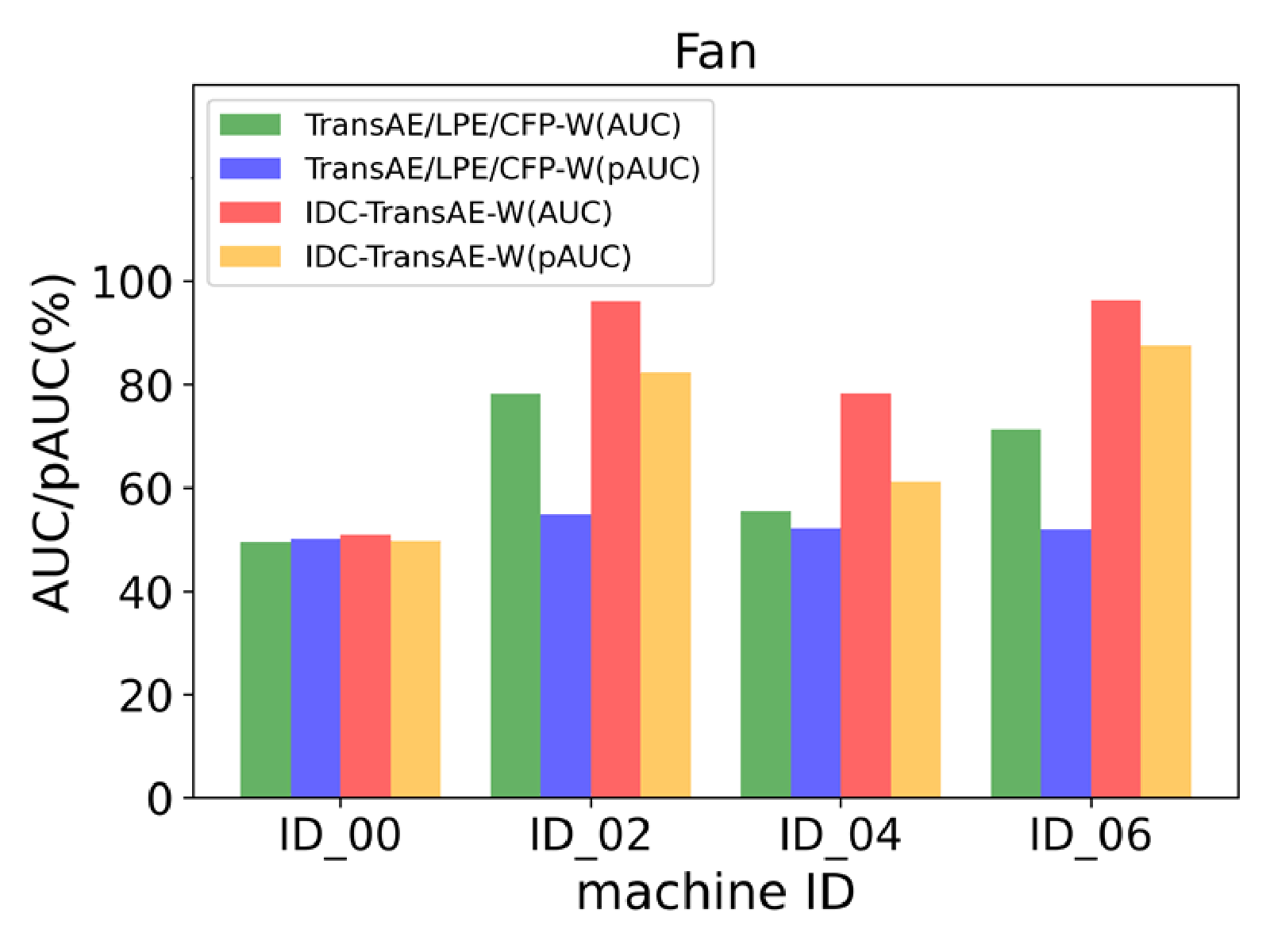}}
	\caption{Performance illustration for 4 different machines with the same type, i.e., Fan.}
	\label{fig:6}
\end{figure}

To further demonstrate the effectiveness of the IDC module, we compare the performance of IDC-TransAE-W and TransAE/LPE/CFP-W in terms of AUC and pAUC on four different machines of the machine type Fan. The result is illustrated in Figure~\ref{fig:6}. From Figure~\ref{fig:6}, we can see that IDC-TransAE-W can significantly improve the performance on ID\_02, ID\_04 and ID\_06, as compared with TransAE/LPE/CFP-W. This means the IDC method can better distinguish the anomalous sound for different machines with the same type. The results in Table~\ref{tab:4} and Figure~\ref{fig:6} verify the effectiveness of different modules of our proposed method. To further illustrate the effectiveness of each module, we give the visualization analysis for each module in the following Section \ref{sec:visual}.

\subsection{Visualization Analysis}
\label{sec:visual}

In this section, visualization analysis is provided for better understanding the experimental results in the ablation studies. Specifically, the effectiveness of CFP, LPE module and IDC module in our proposed IDC-TransAE method are further evaluated. Besides, the influence of the parameter in the GWRP operation of anomaly score calculation is also explored in this section.

\subsubsection{Effectiveness of CFP}
\begin{figure}[ht]
    \centering
    \subfloat[TransAE/PE-W]{
    \label{fig:7a}
    \includegraphics[width=0.495\textwidth]{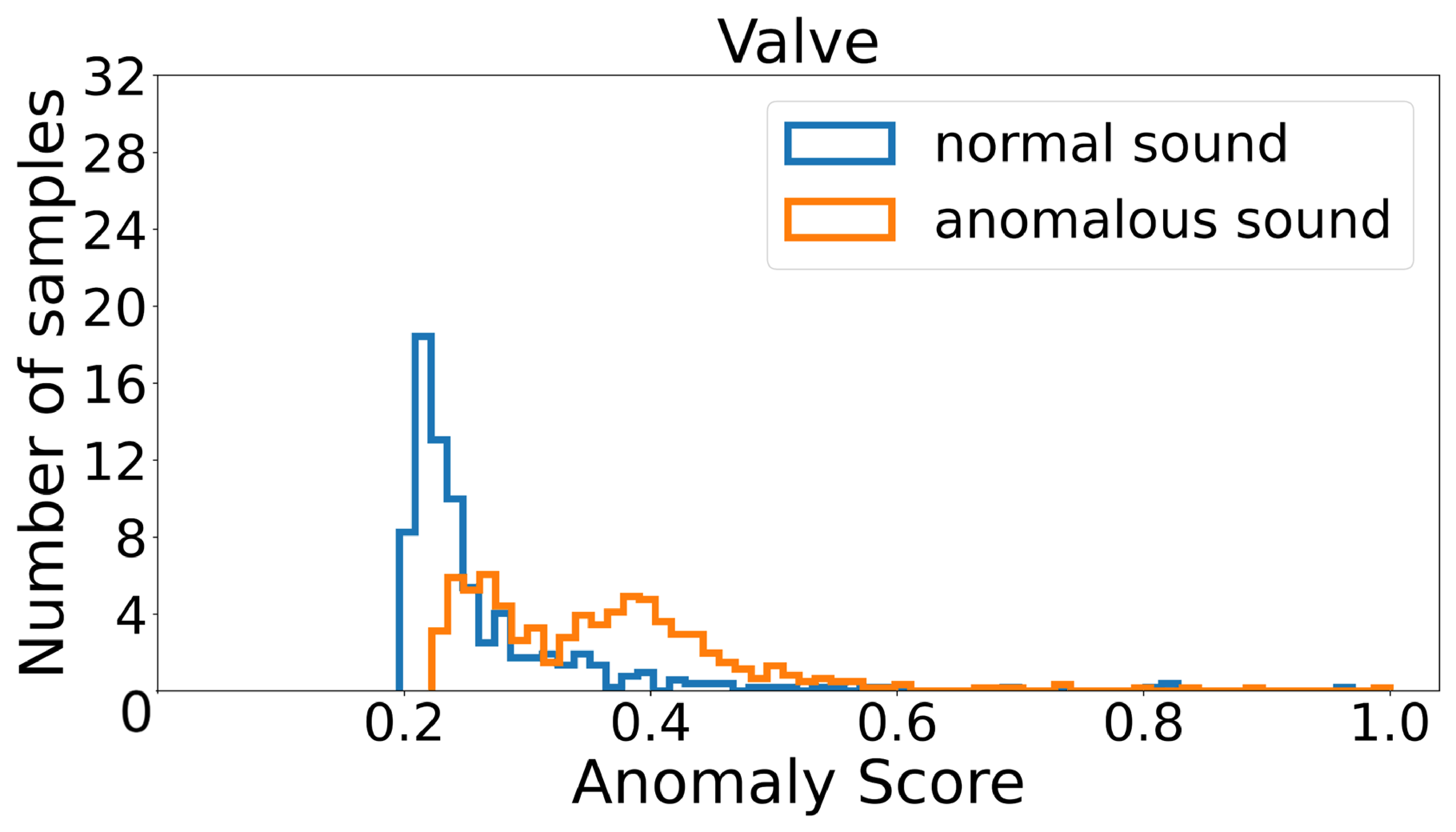}
    }
    \subfloat[TransAE/PE/CFP-W]{
    \label{fig:7b}
    \includegraphics[width=0.495\textwidth]{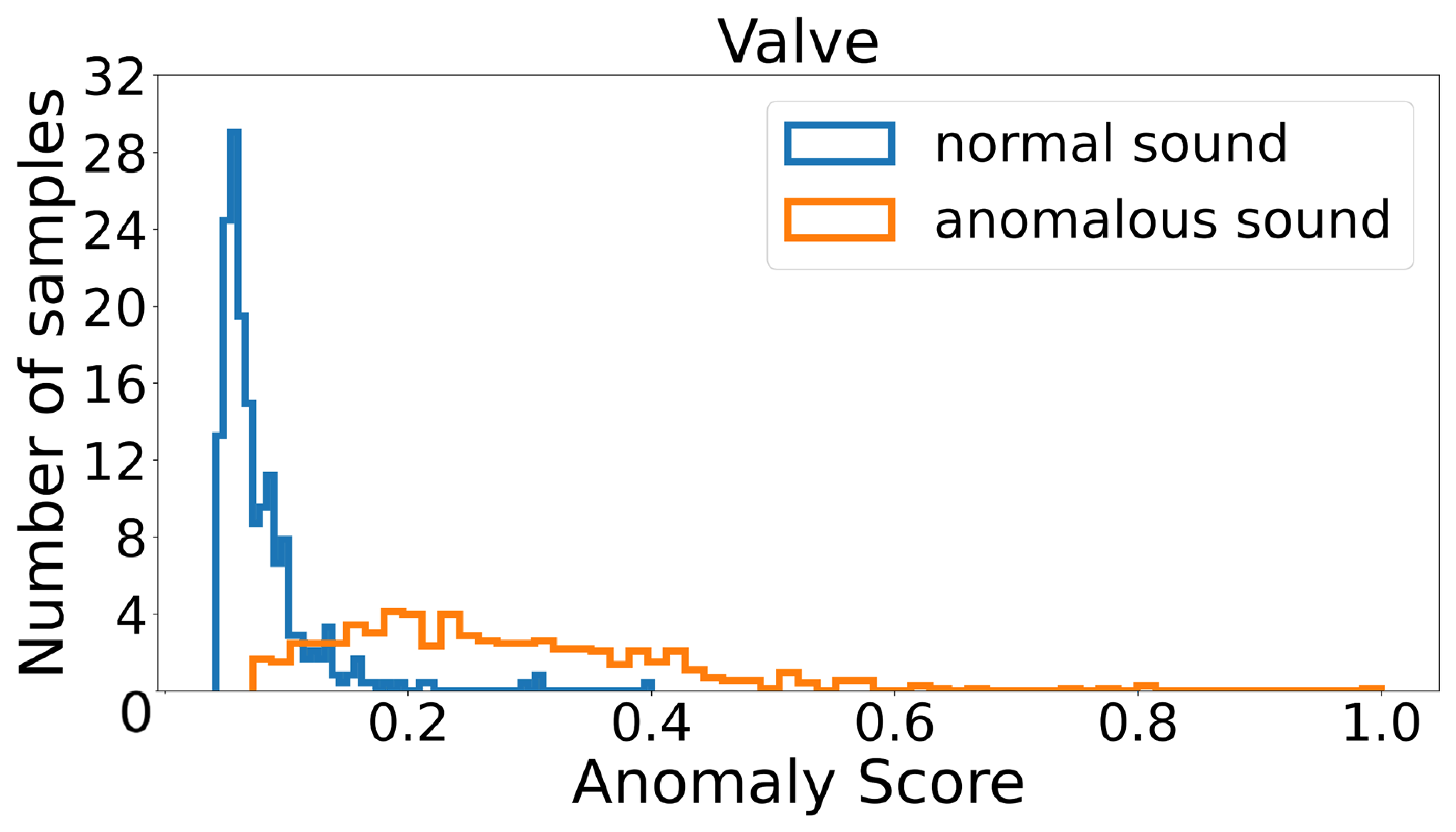}
    }
    \caption{The comparison between TransAE/PE-W and TransAE/PE/CFP-W on histograms of anomaly scores distribution of Valve.} 
\label{fig:7}
\end{figure}

To show how CFP operation affects the anomaly detection for non-stationary sound signals, we compare the histograms of anomaly score distribution between TransAE/PE-W and TransAE/PE/CFP-W on Valve. The result is given in Figure~\ref{fig:7}. Same as Figure~\ref{fig:4}, the anomaly score is also normalized to facilitate comparison. By comparing Figure~\ref{fig:7}(a) and Figure~\ref{fig:7}(b), we can see that the distribution of normal sound samples is on a smaller range of anomaly scores when adopting the CFP module (i.e., TransAE/PE/CFP-W), as illustrated in Figure~\ref{fig:7}(b). It verifies that CFP operation can improve the reconstruction of non-stationary signals as described in~\cite{suefusa2020anomalous}. Therefore, it can improve the performance of our proposed method for anomaly detection of non-stationary sound signals.
\subsubsection{Visualization of Linear Phase Embedding}
\begin{figure}[htbp]
    \centering
    \subfloat[PE]{
        \label{fig:8a}
        \includegraphics[width=0.3\textwidth,height=0.4\textheight]{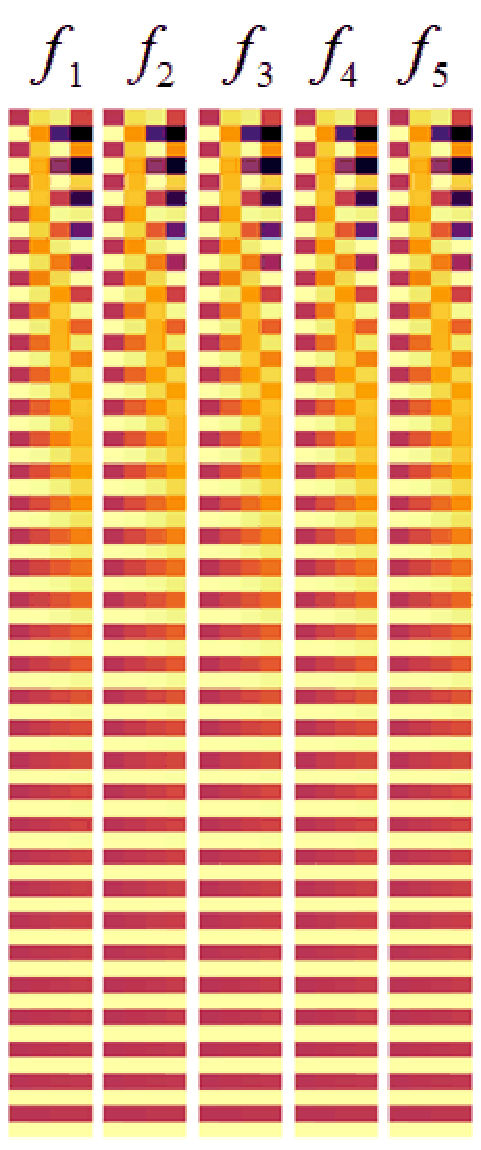}}
    \subfloat[LPE]{
        \label{fig:8b}
        \includegraphics[width=0.3\textwidth,height=0.4\textheight]{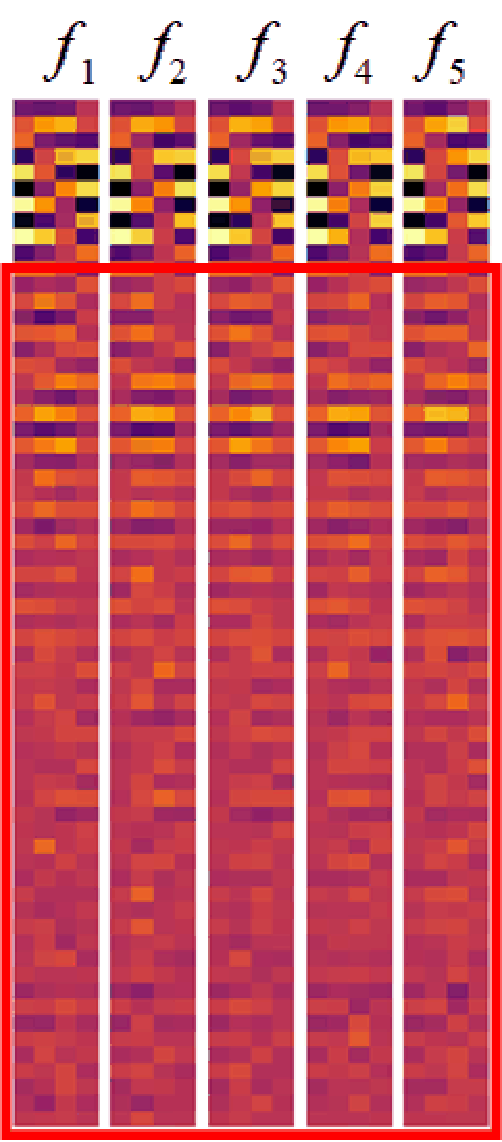}
    }
    \caption{Encoding visualization of five consecutive input sound signals using Positional Encoding (PE) and Linear Phase Embedding (LPE), respectively.} 
\label{fig:8}
\end{figure}

To show why the LPE module can enhance the ability of the model for anomalous sound detection, we visualize the encoding result of five consecutive input sound signals of TransAE/LPE/CFP, and compare it with the encoding result of positional encoding for TransAE/PE/CFP, as illustrated in Figure~\ref{fig:8}. Here, $f_1$ to $f_5$ are the encoding visualizations corresponding to the five consecutive input sound signals, respectively, where each input includes four frames. 

From Figure~\ref{fig:8}(a), we can see that the positional encoding visualization of each input is the same because the positional encoding operation adopts the same cosine representation for signal encoding. In contrast, by linearly encoding the phase information of the signal, our proposed LPE can preserve the signal's own temporal information and give different encoding representations for each different input signal, as indicated in the red box in Figure~\ref{fig:8}(b). Therefore, our proposed method can learn better latent features with unique characteristics from each signal, and enhance the ability of the model for anomalous sound detection.
\subsubsection{Validation of IDC Module}
\begin{figure}[!h]
    \centering
    \subfloat[TransAE/LPE/CFP]{
    \label{fig:9a}
    \includegraphics[width=0.485\textwidth]{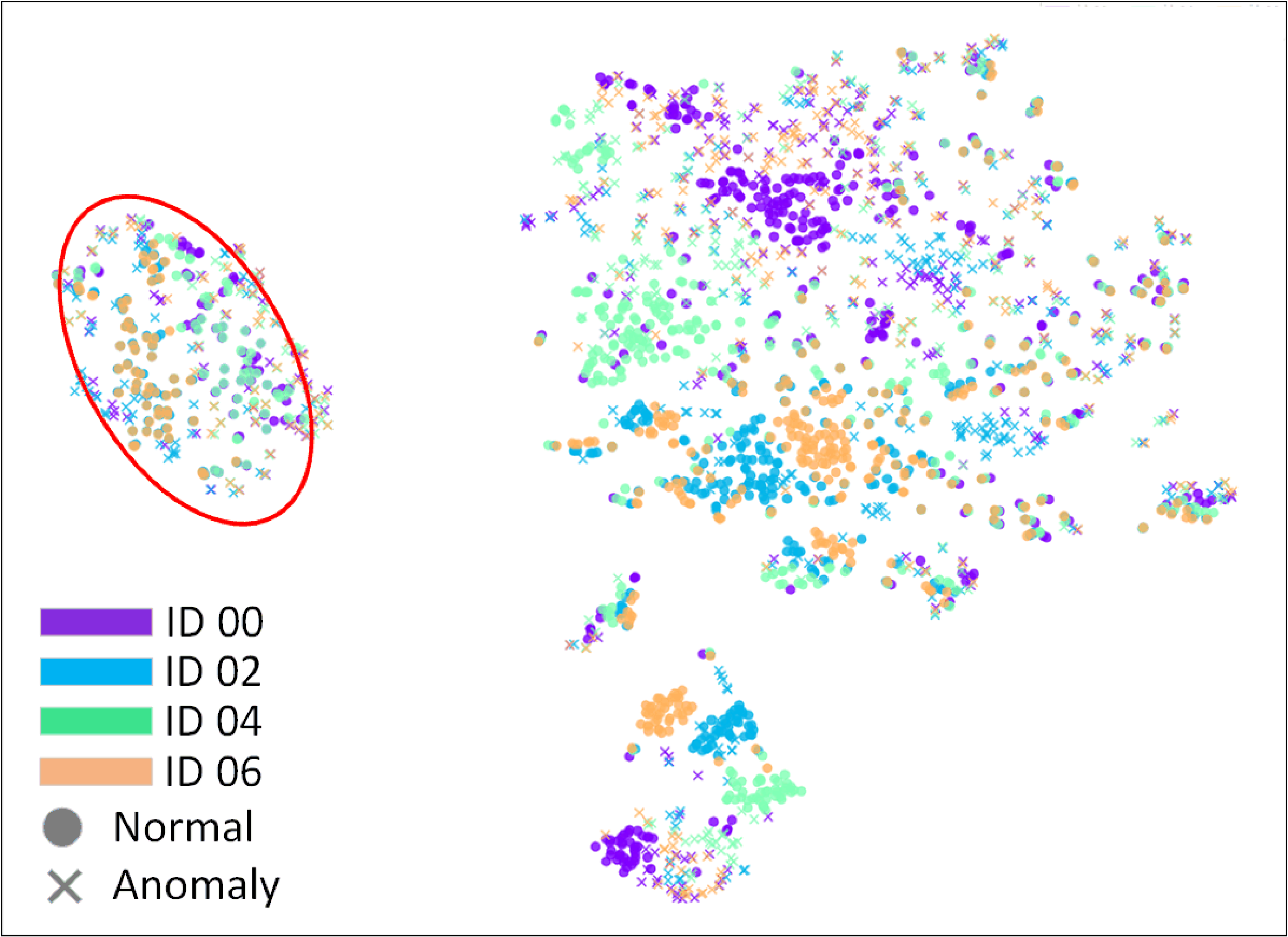}
    }
    \subfloat[IDC-TransAE]{
    \label{fig:9b}
    \includegraphics[width=0.485\textwidth]{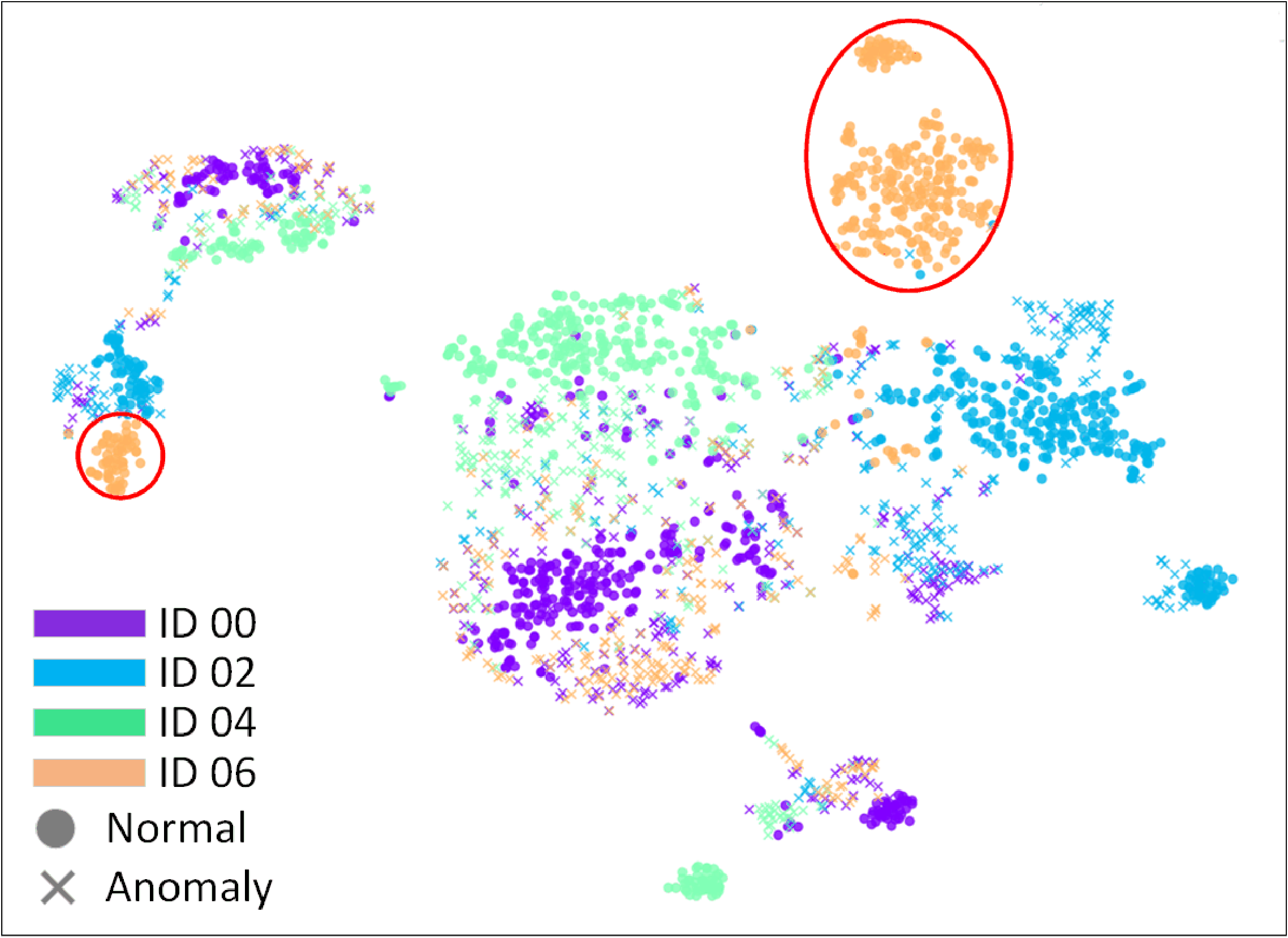}
    }
    \caption{The t-SNE visualization of latent feature on the test dataset for the machine type ToyCar using TransAE/LPE/CFP and IDC-TransAE. Different color represents different machine ID. The ``$\bullet$'' and ``$\times$'' denote normal and anomalous samples, respectively. }
\label{fig:9}
\end{figure}

We show the t-distributed stochastic neighbor embedding (t-SNE) cluster visualization of the latent features to validate the IDC module further. The experiment is conducted on the test dataset of the machine type ToyCar, where the proposed method IDC-TransAE without using ID information (i.e., TransAE/LPE/CFP) is employed for comparison. The result is illustrated in Figure~\ref{fig:9}. 

As observed from Figure~\ref{fig:9}(a),  the latent features of normal and anomalous sound samples from different machines overlap with each other when using the method without IDC module (i.e., TransAE/LPE/CFP). In addition, the latent features of the normal sound samples of one machine may be close to that of the anomalous samples from other machines, rather than the normal samples from the same machine, as illustrated in Figure~\ref{fig:9}(a). It results in the latent features of some anomalous samples from one machine on the manifold of the normal samples from another machine. Thereby these anomalous sounds will be well reconstructed, making it hard to distinguish the anomalies and reducing the detection performance. By introducing the IDC module to constrain the latent feature, the proposed method can reduce the generalization of AE for anomalous sound and further improve its distinguishing ability that the normal and anomalous latent features are well separated, as illustrated in Figure~\ref{fig:9}(b). 

\subsubsection{Influence of Parameter $r$ for Anomaly Detection}
\begin{figure}[!h]
  \centering
    \includegraphics[width=0.85 \textwidth]{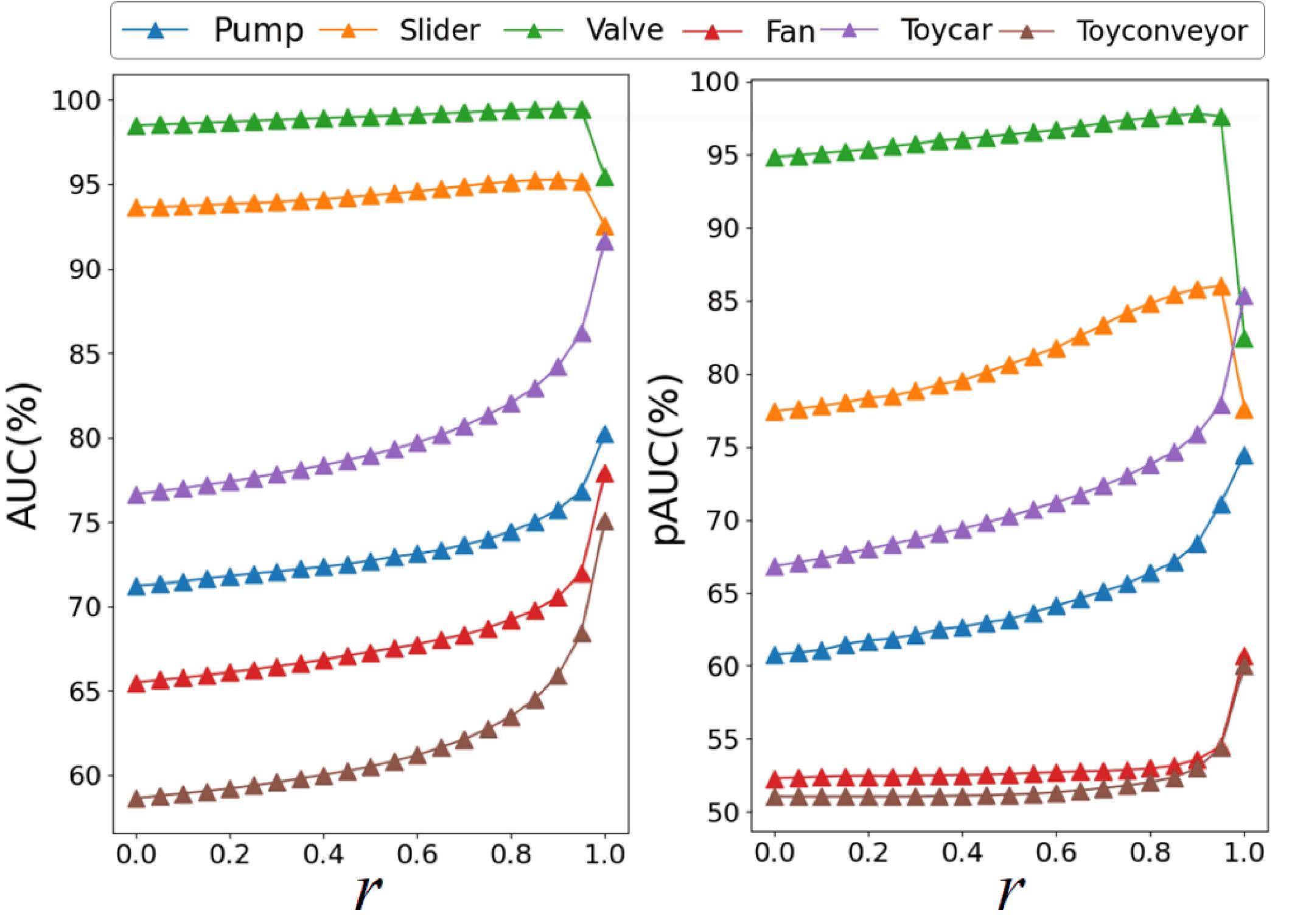}
	\caption{Performance of our proposed IDC-TransAE in terms of AUC and pAUC  under different $r$ values for all six machine types.}
	\label{fig:10}
\end{figure}

%
As mentioned in Section \ref{sec:3_2}, we introduce the weighted anomaly score computation to highlight the anomalous events that only appear for a short time. The parameter $r$ in Equation \eqref{eq:10} will decide the way for anomaly score computation, i.e., weighted anomaly score computation will degenerate to max anomaly score computation when $r=0$, and it will become mean anomaly score computation when $r=1$. Therefore, we also carry out another experiment to show the impact of parameter $r$ on the performance of our proposed IDC-TransAE for anomaly detection. Here, different values of $r$  from  $0 \leq r \leq 1$ with an interval of $0.05$ are selected to evaluate the performance of our proposed method in terms AUC and pAUC on all six machine types. The result is shown in Figure~\ref{fig:10}.

From Figure~\ref{fig:10}, we can see that the mean score computation (i.e., $r=1$) can achieve the best performance for the machine types of Fan, Pump, ToyCar and ToyConveyor. However, it obtains the worst performance for the machine types of Slider and Valve, where the anomalous sound often occurs in a short time. Though using max anomaly score computation ($r=0$) can achieve better performance than adopting mean anomaly score computation on Slider and Valve, the weighted anomaly score computation method can provide the best performance for the machine type of Slider and Valve. Especially, the weighted anomaly score computation method can significantly improve the pAUC performance over the mean and max anomaly score computation on Slider and Valve. The result verifies the effectiveness of  weighted anomaly score computation for the anomalous sound that appears over short time. In addition, the values of $r$ can be adjusted according to different machine types, which makes it more applicable than mean anomaly score and max anomaly score computation.
\section{Conclusions}
\label{sec:5}
In this paper, we have presented an IDC-TransAE architecture with weighted anomaly score computation for unsupervised ASD, where an ID classifier was introduced to mitigate the generalization of AE for anomalous sound and enhance the distinguishing ability for different machines with the same type. In addition, center frame prediction was utilized to improve the reconstruction of the non-stationary sound signal, and a linear phase embedding strategy was applied to preserve the signal's temporal information and further improve its distinguishing ability for anomalous sound detection. Moreover, a weighted anomaly score computation method was introduced to highlight the anomaly scores for anomalous events that only appear for a short time. The experiments demonstrate the effectiveness and superiority of our proposed method, as compared with the baseline methods.

\section{Declarations}

\textbf{Availability of data and materials}\\
The datasets for experimental evaluation in this study are from the DCASE 2020 challenge Task 2, which are available on the internet.\\ \hspace*{\fill}

\noindent \textbf{Funding}\\
This work was partly supported by the Natural Science Foundation of Heilongjiang Province under Grant No. YQ2020F010, and a GHfund with Grant No. 202302026860.\\ \hspace*{\fill}

\noindent \textbf{Acknowledgements}\\
The authors would like to thank the Associate Editor and the anonymous reviewers for reviewing the manuscript.

\bibliography{mybibfile}

\end{document}